\documentclass[10pt, journal]{IEEEtran}

\usepackage[usenames, x11names*]{xcolor}
\usepackage[T1]{fontenc}
\usepackage[utf8]{inputenc}

\usepackage[english]{babel}
\usepackage{csquotes}
\usepackage{mathtools,amssymb,amsthm}
\usepackage{bm}
\usepackage{nicefrac}
\usepackage{subcaption}
\usepackage{booktabs}
\usepackage{float}
\usepackage{url}
\usepackage[bookmarks=false]{hyperref}
\usepackage{tikz}
\usepackage{pgfplots}
\usetikzlibrary{quantikz}
\usepackage{graphicx}
\usepackage{empheq}
\usepackage[binary-units=true, decimalsymbol=comma, output-decimal-marker={.}]{siunitx}
\usepackage[textsize=tiny]{todonotes}
\usetikzlibrary{arrows,backgrounds,calc,chains,matrix,positioning,shapes,shapes.geometric,	shapes.arrows, decorations.pathmorphing, decorations.pathreplacing}
\tikzset{%
	font={\footnotesize},
	vertex/.style={draw,circle,inner sep=0pt,minimum width=0.5cm,minimum height=0.5cm},
	zeroterm/.style={below,inner sep=0pt,font=\tiny}
}
\pgfplotsset{compat=newest}

\usepackage{csvsimple}

\usepackage{stmaryrd}

\usepackage{nicematrix}
\usepackage{ifthen}
\usepackage{etoolbox}

\makeatletter
\def\thm@space@setup{%
  \thm@preskip=.7\topsep
  \thm@postskip=\thm@preskip %
}
\makeatother

\newtheorem{example}{Example}
\newcommand{\qop}[1]{\ensuremath{\mathit{#1}}}

\title{\resizebox{\linewidth}{!}{Advanced Equivalence Checking for Quantum Circuits}}

\author{
Lukas Burgholzer\IEEEauthorrefmark{1}~\IEEEmembership{Student Member,~IEEE}\hspace{3cm}Robert Wille\IEEEauthorrefmark{1}\IEEEauthorrefmark{2}~\IEEEmembership{Senior Member,~IEEE}\\
\IEEEauthorrefmark{1}Institute for Integrated Circuits, Johannes Kepler University Linz, Austria\\
\IEEEauthorrefmark{2}Software Competence Center Hagenberg GmbH (SCCH), Austria\\
lukas.burgholzer@jku.at\hspace{4cm}robert.wille@jku.at\vspace*{-2.5em}}

\hypersetup{
	pdftitle={Advanced Equivalence Checking for Quantum Circuits},
	pdfsubject={IEEE Trans. on CAD of Integrated Circuits and Systems},
	pdfauthor={Lukas Burgholzer and Robert Wille}
}

\usepackage[style=ieee, maxnames=5, minnames=1, date=year, doi=false,isbn=false,url=false,backend=biber, sortcites, dashed=false]{biblatex}
\usepackage[keeplastbox]{flushend}
\begin{document}\maketitle

\begin{abstract}
In the not-so-distant future, quantum computing will change %
the way we tackle certain problems. It promises to dramatically speed-up many chemical, financial, cryptographical, and machine-learning applications. However, in order to capitalize on those promises, complex design flows composed of steps such as compilation, decomposition, mapping, or transpilation need to be employed before being able to execute a conceptual quantum algorithm on an actual device. This results in many descriptions at various levels of abstraction which may significantly differ from each other. %
The complexity of the underlying design problems makes it ever more important to not only provide efficient solutions for the single steps, but also to verify that the originally intended functionality is %
preserved throughout all levels of abstraction. This motivates methods for equivalence checking of quantum circuits.
However, most existing methods for this are inspired by equivalence checking in the classical realm and have merely been extended to support quantum circuits (i.e.,~circuits which do not only rely on 0's and 1's, but also employ superposition and entanglement). 

In this work, we propose an advanced methodology which takes the different paradigms of quantum circuits not only as a burden, but as an opportunity.  In fact, the proposed methodology explicitly utilizes characteristics unique to quantum computing in order to overcome the shortcomings of existing approaches. 
We show that, by exploiting the reversibility of quantum circuits, complexity can be kept feasible in many cases. 
Moreover, we show that, in contrast to the classical realm, simulation is very powerful in verifying quantum circuits.
Experimental evaluations confirm that the resulting %
methodology allows one to conduct equivalence checking %
dramatically faster than ever before---in many cases just a single simulation run is sufficient.
An implementation of the proposed equivalence checking flow is publicly available at \url{https://iic.jku.at/eda/research/quantum_verification/}.
\end{abstract}

\vspace*{-2mm}
\section{Introduction}\label{sec:intro}
\vspace*{-1mm}

Quantum computers~\cite{nielsenQuantumComputationQuantum2010} aim to change the way we tackle certain problems in the future. As more and more big companies like Google, IBM, Intel, and Microsoft as well as \mbox{start-ups} like Rigetti and IonQ set foot in this domain, the need for design methods that allow to use this new technology is steadily increasing. Without appropriate methods, we might reach a point where we have quantum computers readily available but no means to actually use them.
In order to utilize the theoretical advantage of quantum computers in practice, a multitude of (computationally complex) design tasks have to be conducted---resulting in descriptions of quantum algorithms at various abstraction levels which may significantly differ %
in their basis operations and structure. This is similar to the classical realm where, e.g., descriptions at the \emph{Electronic System Level}, the \emph{Register Transfer Level}, and the \emph{Gate Level} exist.

More precisely, %
a high-level description of a quantum algorithm has to be \emph{compiled} to a low-level description satisfying all constraints imposed by the targeted device. Today's quantum computers
only support a very limited (yet universal) set of quantum operations natively. Thus, non-native operations first have to be \emph{decomposed} into sequences of native operations~\cite{gilesExactSynthesisMultiqubit2013, amyMeetinthemiddleAlgorithmFast2013, willeImprovingMappingReversible2013, millerElementaryQuantumGate2011, maslovAdvantagesUsingRelative2016}.  %
Furthermore, most quantum computers limit the pairs of qubits (the main computational unit in quantum computing) that may directly interact with each other. Realizing a generic quantum algorithm on such a device therefore requires a \emph{mapping} step (sometimes also called \emph{transpilation} or \emph{qubit routing}), where a given circuit is made compliant to the imposed connectivity constraints by inserting %
\qop{SWAP} or \qop{H}~gates~\cite{siraichiQubitAllocation2018,zulehnerEfficientMethodologyMapping2019, zulehnerCompilingSUQuantum2019, smithQuantumComputationalCompiler2019, willeMappingQuantumCircuits2019, matsuoEfficientMethodQuantum2019, amyStaqFullstackQuantum2020, cowtanQubitRoutingProblem2019}. 
Finally, current devices are heavily affected by noise and decoherence effects and are considered part of the \emph{Noisy-Intermediate-Scale-Quantum}~(NISQ) era of quantum computing~\cite{preskillQuantumComputingNISQ2018}.
On the one hand, this motivates quantum circuit optimizations, such as gate fusion, gate cancellation, or block-wise re-synthesis---which aim to reduce the overall gate count of circuits to be executed in order to reduce the effect of noise and allow the computation to stay coherent~\cite{hattoriQuantumCircuitOptimization2018, sasanianReversibleQuantumCircuit2013, vidalUniversalQuantumCircuit2004,itokoQuantumCircuitCompilers2019, namAutomatedOptimizationLarge2018, maslovQuantumCircuitSimplification2008, prasadDataStructuresAlgorithms2006, iwamaTransformationRulesDesigning2002,duncanGraphtheoreticSimplificationQuantum2019}.
On the other hand, mapping techniques which take the targeted device's calibration and error data into account to achieve noise-adaptive mappings shift into focus~\cite{bhattacharjeeMUQUTMulticonstraintQuantum2019, muraliNoiseadaptiveCompilerMappings2019}. 

All this substantially changes the circuit description during the design process.
At the same time, all these steps should obviously preserve the originally intended functionality of the quantum circuit---even if other basic operations and structures are eventually used for the realization.
In order to check that, \emph{equivalence checking} is usually conducted to prove whether two circuits (the originally given quantum algorithm and the quantum circuit resulting from the compilation process) are equivalent or to determine a counterexample showing the non-equivalence between them. 
In the classical realm, %
equivalence checking is conducted using design automation expertise leading to efficient methods in order to guarantee correctness throughout the design~\cite{brandVerificationLargeSynthesized1993, molitorEquivalenceCheckingDigital2010, marques-silvaCombinationalEquivalenceChecking1999, jhaEquivalenceCheckingUsing1997}.

Inspired by these methods, several solutions for equivalence checking of quantum circuits have been proposed in the recent past, e.g., based on
re-writing~\cite{yamashitaFastEquivalencecheckingQuantum2010, duncanGraphtheoreticSimplificationQuantum2019}, 
Boolean satisfiability~\cite{willeCompactEfficientSAT2013}, and
decision diagrams~\cite{wangXQDDbasedVerificationMethod2008, niemannQMDDsEfficientQuantum2016,viamontesCheckingEquivalenceQuantum2007,niemannEquivalenceCheckingMultilevel2014}.
However, all of them merely extended classical methods in order to additionally support quantum circuits (i.e., are extended to support superposition or entanglement) and, by this, take the different computation paradigm only as a burden to be addressed. Consequently, %
those methods remain unsatisfactory in many cases.

In this work\footnote{Preliminary versions of this work have been published in~\cite{burgholzerImprovedDDbasedEquivalence2020, burgholzerPowerSimulationEquivalence2020}.}, we propose a different take. Rather than a burden, we see the computation paradigm of quantum circuits %
as an opportunity. We propose an advanced equivalence checking methodology which explicitly utilizes characteristics unique to quantum computing in order to substantially improve existing approaches.
More precisely, we unearth potential which rests on the following two observations:
\begin{itemize}
\item Quantum circuits are inherently reversible. Because of that, if two quantum circuits $G$ and $G^\prime$ are equivalent, then concatenating the first circuit~$G$ with the inverse~$G^{\prime -1}$ of the second circuit would realize the identity function~$\mathbb{I}$, i.e.,  $G \cdot G^{\prime -1} = \mathbb{I}$. The potential now lies in the order in which the operations from either circuit are applied. Whenever a strategy can be employed so that the respective gates from $G$ and $G^{\prime}$ are applied in a fashion frequently yielding the identity, the entire procedure can be conducted rather efficiently since the identity constitutes the best case for most representations of quantum functionality (e.g., linear in the number of qubits for decision diagrams). 
	
	\item Moreover, even in the case where the two considered quantum circuits are not equivalent, quantum characteristics can be exploited. In fact, 
	we observed that, again due to the inherent reversibility of quantum operations, even small differences in quantum circuits frequently affect the \emph{entire} functional representation.  Hence, it may not always be
	necessary to check the complete functionality, but it is highly likely that the simulation of both computations with a couple of arbitrary input states (i.e.,~considering only a small part of the whole functionality) will already provide a counterexample showing the non-equivalence. This is in stark contrast to the classical realm, where the inevitable information loss introduced by many logic gates and the resulting masking effects often require a complete consideration of \emph{all} possible input states or sophisticated schemes for constraint-based stimuli generation~\cite{yuanConstraintbasedVerification2006,bergeronWritingTestbenchesUsing2006,kitchenStimulusGenerationConstrained2007,willeSMTbasedStimuliGeneration2009}, fuzzing~\cite{laeuferRFUZZCoveragedirectedFuzz2018,leDetectionHardwareTrojans2019}, etc. 
\end{itemize}

Both observations provide the nucleus of an advanced equivalence checking methodology in which the non-equivalence is often detected by a few simulation runs (which can be conducted dramatically faster than the actual equivalence check). Moreover, passing several simulation runs leading to the same results for both circuits provides an indication (albeit no proof) that the circuits might be equivalent. The proof itself can, afterwards, be significantly accelerated by using strategies which keep the check for $G \cdot G^{\prime-1}=\mathbb{I}$ close to the identity~$\mathbb{I}$.
Combining these complementary ideas into a comprehensive equivalence checking flow allows for efficient verification of quantum circuits.

Experimental evaluations confirm that the resulting flow allows one to conduct equivalence checking %
dramatically faster than ever before---in many cases, just a single simulation run is sufficient. 
By this, we do not only show ways to handle the complexity of verifying quantum circuits, but also show the potential of 
simulation for this task.
An implementation of the proposed equivalence checking flow is publicly available at \url{https://iic.jku.at/eda/research/quantum_verification/}. %

The remainder of this paper is structured as follows: Section~\ref{sec:background} reviews the %
background needed to keep this work \mbox{self-contained}.
Section~\ref{sec:motivation} motivates the considered problem and discusses the related work.
Then, Section~\ref{sec:general} introduces and illustrates the general ideas proposed %
in this work.
Based on that, Section~\ref{sec:strategies} describes the dedicated equivalence checking schemes resulting from the ideas and provides a theoretical discussion
on the power of simulation for equivalence checking of quantum circuits.
All these findings eventually result in an advanced equivalence checking methodology which is described and discussed in Section~\ref{sec:ecflow}.
Finally,  
Section~\ref{sec:experiments} summarizes the obtained experimental results before %
Section~\ref{sec:conclusion} concludes the paper.

\vspace*{-2mm}
\section{Background}\label{sec:background}\vspace*{-1mm}
In this section, we review %
the main concepts of quantum computing and illustrate decision diagrams as one way of efficiently representing and manipulating quantum functionality. While the following descriptions are kept brief, we refer to~\cite{nielsenQuantumComputationQuantum2010} and~\cite{niemannQMDDsEfficientQuantum2016} for more details on either topic.

\vspace*{-2mm}
\subsection{Quantum Computing}\label{sec:qc}

The main computational unit in quantum computing is the \emph{qubit}. In contrast to classical bits, a qubit cannot only be in one of the computational basis states $\ket{0}$ or $\ket{1}$ (written in Dirac notation), but also in an arbitrary \emph{superposition} of these states. Specifically, the state $\ket{\varphi}$ of a qubit is described by two \emph{amplitudes} $\alpha_0,\alpha_1\in\mathbb{C}$ such that 
\[
\ket{\varphi} = \alpha_0 \ket{0} + \alpha_1 \ket{1} \equiv
\alpha_0 \begin{bNiceMatrix}[small] 1\\0 \end{bNiceMatrix} + 
\alpha_1 \begin{bNiceMatrix}[small] 0\\1 \end{bNiceMatrix} =
\begin{bNiceMatrix}[small] \alpha_0\\\alpha_1 \end{bNiceMatrix}
\]
 and $\vert\alpha_0\vert^2+\vert\alpha_1\vert^2=1$. The resulting column vector is also referred to as \emph{state vector}.
In a system of $n$ qubits, there exist $2^n$ computational basis states $\ket{i}$ with $i$ from $0$ to $2^n-1$. Analogously, the state of an $n$-qubit system is described by $2^n$ complex amplitudes $\alpha_i\in\mathbb{C}$ such that $\ket{\varphi} = \sum_{i=0}^{2^n-1} \alpha_i \ket{i}$ and $\sum_{i=0}^{2^n-1} \vert \alpha_i\vert^2=1$---which can again be interpreted as a state vector, i.e., \mbox{$\ket{\varphi} \equiv [\alpha_{0}, \, \dots ,\, \alpha_{2^n-1}]^\top$}.

\begin{example}\label{ex:state}
	Consider a two qubit system whose state is described by the state vector \[
	\tfrac{1}{\sqrt{2}}[1,\,0,\,0,\,1]^\top \equiv \tfrac{1}{\sqrt{2}} \ket{0} + \tfrac{1}{\sqrt{2}} \ket{3} = \tfrac{1}{\sqrt{2}} \ket{(00)_2} + \tfrac{1}{\sqrt{2}} \ket{(11)_2}.
	\]
	This is a valid quantum state, since \mbox{$\vert\nicefrac{1}{\sqrt{2}}\vert^2 + \vert\nicefrac{1}{\sqrt{2}}\vert^2 = 1$}. Furthermore, it demonstrates a key phenomenon unique to quantum computing called \emph{entanglement}.
	While the complete system's state can be accurately described (by the above statevector), its individual parts, i.e., the state of the individual qubits, cannot.
	More precisely, the state $\ket{\varphi}$ cannot be split into $\ket{q_1}\otimes\ket{q_0}$, where $\ket{q_i}$ describes the state of the $i^{\mathit{th}}$ qubit and $\otimes$ denotes the tensor product of both vectors.
\end{example}

In a real quantum system, the individual amplitudes $\alpha_i$ are not directly observable. Instead, a \emph{measurement} operation collapses the system's state to one of the computational basis states $\ket{i}$---each with probability $\vert\alpha_i\vert^2$---which can then be read-out classically.

\begin{example}\label{ex:measure}
	The state $\nicefrac{1}{\sqrt{2}} \ket{(00)_2} + \nicefrac{1}{\sqrt{2}} \ket{(11)_2}$ from Ex.~\ref{ex:state} represents an equal superposition of two basis states.
	Measuring this state would yield $\ket{(00)_2}$ in half of the cases and~$\ket{(11)_2}$ otherwise.
\end{example}

The state of a quantum system is manipulated by \emph{quantum operations} (often also referred to as \emph{quantum gates}). 
Typically, these operations only operate on a small subset of a system's qubits.
An operation acting on $k\leq n$ qubits (most frequently \mbox{$k=1$} or \mbox{$k=2$}) is described by a $2^k\times 2^k$ unitary matrix\footnote{A complex matrix $U$ is unitary if $U^\dag U = U U^\dag = \mathbb{I}$, where $U^\dag$ denotes the conjugate-transpose of $U$ and $\mathbb{I}$ the identity matrix.}~$U$. Applying such an operation to an $n$-qubit system in the state~$\ket{\varphi}$ corresponds to extending the (local) $2^k \times 2^k$ matrix to a $2^n\times 2^n$ \emph{(system) matrix} using tensor products and, then, calculating the matrix-vector product~\mbox{$U \ket{\varphi}$}---resulting in a new state~$\ket{\varphi^\prime}$\footnote{As a consequence, quantum operations are inherently reversible,~i.e., $U^\dag \ket{\varphi^\prime} = U^\dag U \ket{\varphi} = \ket{\varphi}$.}.

\begin{example}
	Consider a single-qubit operation (i.e., $k=1$) represented by the unitary matrix $U_s\in\mathbb{C}^{2\times 2}$.
	Further assume that this operation is to be applied to the second qubit in a system consisting of $n=3$ qubits.
	Then, the full $2^3\times 2^3$ matrix corresponding to this operation is given by $\mathbb{I}_2\otimes U_s \otimes \mathbb{I}_2$, where $\mathbb{I}_2$ denotes the $2\times 2$ identity matrix. Its application to the \mbox{$2^3$-dimensional} state vector $\ket{\varphi} \equiv [\alpha_{0}, \, \dots ,\, \alpha_{7}]^\top$ then yields the new state $\ket{\varphi^\prime}$.
\end{example}

\emph{Quantum computations} are just sequences of quantum operations applied to the state of a system.
This is predominantly described by \emph{quantum circuits} and visualized as \emph{quantum circuit diagrams}.
There, horizontal wires indicate the system's individual qubits, while gates that are placed on these wires indicate the sequence of operations to apply---operating from left to right.

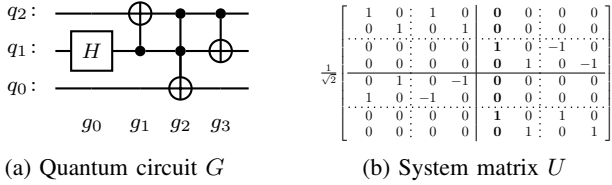
\begin{figure}[t]
	\centering
	\begin{subfigure}[b]{0.45\linewidth}
		\centering
		\resizebox{0.9\linewidth}{!}{
			\begin{quantikz}[column sep=6pt, row sep={0.5cm,between origins}, ampersand replacement=\&]
				\lstick{$q_2 \colon$} \& \qw \& \targ{} \& \ctrl{2} \& \ctrl{1} \& \qw \\
				\lstick{$q_1\colon$}  \& \gate{H} \& \ctrl{-1} \& \ctrl{1} \&\targ{}\& \qw \\
				\lstick{$q_0\colon$} \& \qw  \& \qw \& \targ{} \& \qw \& \qw \\
				\& \lstick[label style={xshift=0.3cm}]{$g_0$} \& \lstick[label style={xshift=0.3cm}]{$g_1$} \& \lstick[label style={xshift=0.3cm}]{$g_2$} \& \lstick[label style={xshift=0.3cm}]{$g_3$} 
		\end{quantikz}}
		\caption{Quantum circuit $G$}
		\label{fig:g}
		\vspace*{-1mm}
	\end{subfigure}%
	\hfill
	\begin{subfigure}[b]{0.5\linewidth}
		\centering
		\resizebox{0.9\linewidth}{!}{
			$\frac{1}{\sqrt{2}}\begin{bNiceArray}{RR:RR|RR:RR}[create-medium-nodes, margin, columns-width = auto]
			1 & 0 & 1 & 0 & \bm{0} & 0 & 0 & 0 \\
			0 & 1 & 0 & 1 & \bm{0} & 0 & 0 & 0 \\\hdottedline
			0 & 0 & 0 & 0 & \bm{1} & 0 & -1 & 0 \\
			0 & 0 & 0 & 0 & \bm{0} & 1 & 0 & -1\\\hline
			0 & 1 & 0 & -1 & \bm{0} & 0 & 0 & 0 \\
			1 & 0 & -1 & 0 & \bm{0} & 0 & 0 & 0 \\\hdottedline
			0 & 0 & 0 & 0 & \bm{1} & 0 & 1 & 0 \\
			0 & 0 & 0 & 0 & \bm{0} & 1 & 0 & 1\\
			\end{bNiceArray}$}
		\caption{System matrix $U$}
		\label{fig:u}
		\vspace*{-1mm}
	\end{subfigure}
	\caption{Quantum computations}
	\label{fig:qc}
	\vspace*{-7mm}
\end{figure}

\begin{example}
	Fig.~\ref{fig:g} shows a quantum circuit~$G$ comprised of three qubits~$q_0,q_1,q_2$ and four gates~$g_0,g_1,g_2,g_3$. 
	The first gate represents a \emph{Hadamard} operation (indicated by a box with the label $H$), which can be used to create an equal superposition from a given basis state (cf.~Ex.~\ref{ex:measure}).
	The remaining three gates represent \emph{controlled} operations. A specific operation (in this case a negation of the qubit state) is applied to a target qubit (indicated by $\oplus$) if and only if certain control qubits (indicated by $\bullet$) are in state $\ket{1}$.
\end{example}
	
Each gate $g_i$ represents a corresponding unitary matrix~$U_i$ that is subsequently applied during the execution of a quantum circuit.
Thus, the functionality of a given circuit \mbox{$G=g_0\dots g_{m-1}$} can be obtained as a unitary \emph{system matrix} $U$ itself by determining \mbox{$U=U_{m-1}\cdots U_0$}. 
Moreover, executing the quantum circuit for a given initial state $\ket{\varphi}$ (also called \emph{simulation} when conducted on a classical computer) leads to an evolution of the state $\ket{\varphi}$ according to \mbox{$U_{m-1}\cdots U_0 \ket{\varphi} = U \ket{\varphi} = \ket{\varphi^\prime}$}.
Note that, if $\ket{\varphi}=\ket{i}$ for some \mbox{$i\in\{0,\dots,2^n-1\}$}, i.e., $\ket{\varphi}$ is a computational basis state, the simulation precisely calculates the $i^{th}$ column $u_i$ of~$U$, i.e.,~\mbox{$U\ket{i}=\ket{u_i}$}.
	
\begin{example}
	Consider again the circuit~$G=g_0g_1g_2g_3$ shown in Fig.~\ref{fig:g}. Consecutively multiplying the corresponding gate matrices \mbox{$U_3\cdot U_2\cdot U_1\cdot U_0$} eventually yields the system matrix~$U$ shown in Fig.~\ref{fig:u}.
	Given the initial state~\mbox{$\ket{4}=\ket{(100)_2}$}, executing the quantum circuit~$G$ precisely transforms this state to the state described by column four of~$U$, i.e.,
	\resizebox{0.99\linewidth}{!}{$
	U \ket{(100)_2} = \ket{u_4} = \nicefrac{1}{\sqrt{2}}\, \begin{bNiceMatrix}[small] 0 & 0 & 1 & 0 & 0 & 0 & 1 & 0\end{bNiceMatrix}^\top = \nicefrac{1}{\sqrt{2}} \ket{(010)_2} + \nicefrac{1}{\sqrt{2}} \ket{(110)_2}$}.
\end{example}

Since quantum operations are inherently reversible, the inverse of a whole quantum circuit~\mbox{$G=g_0\dots g_{m-1}$} can always be determined by reversing the order of operations and inverting each individual operation, i.e.,  
\mbox{$G^{-1} = (g_0\dots g_{m-1})^{-1} = g_{m-1}^{-1}\dots g_0^{-1}$}---corresponding to the matrix operations \mbox{$U_0^{\dag}\cdots U_{m-1}^{\dag}=U^\dag$}.
\vspace*{-2mm}
\subsection{Decision Diagrams for Quantum Computing}\label{sec:dds}

State vectors and operation matrices of a quantum system are exponential in size with respect to the number of qubits---quickly rendering the simulation of a quantum circuit~$G$ and even more the construction of its system matrix~$U$ an extremely difficult task.
\emph{Decision diagrams} (DDs) have been proposed as an efficient way for representing and manipulating quantum functionality~\cite{wangXQDDbasedVerificationMethod2008, millerQMDDDecisionDiagram2006, niemannQMDDsEfficientQuantum2016, zulehnerAdvancedSimulationQuantum2019, zulehnerHowEfficientlyHandle2019}.
While they are still exponential in the worst-case, decision diagrams have been shown to lead to very compact representations for the vectors and matrices in many cases. %
In the following, we review how decision diagrams can be used to compactly represent matrices (e.g., the system matrix $U$). As we will see, (state) vectors are just a special case of matrices that can be handled analogously.

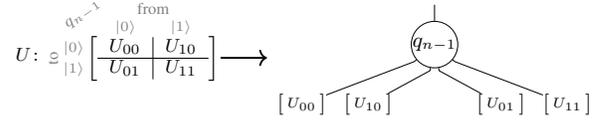
\begin{figure}[t]
	\centering
	\resizebox{0.9\linewidth}{!}{
	\begin{tikzpicture}[vertex/.style={inner sep=0pt,minimum width=0.3cm,minimum height=0.3cm, font=\small}]
	\scalebox{1.0}{
		\matrix[ampersand replacement=\&,every node/.style={vertex},column sep={1.cm,between origins},row sep={.9cm,between origins}] (qmdd) {
			\&\& \node[draw,circle,inner sep=0pt,minimum width=0.7cm,minimum height=0.7cm] (n1) {$q_{n-1}$}; \&\& \\
			\node%
			(a) {$\left[\begin{smallmatrix*}[r] U_{00}\end{smallmatrix*}\right]$}; \&
			\node%
			(b) {$\left[\begin{smallmatrix*}[r] U_{10}\end{smallmatrix*}\right]$}; \& \&
			\node%
			(c) {$\left[\begin{smallmatrix*}[r] U_{01}\end{smallmatrix*}\right]$}; \&
			\node%
			(d) {$\left[\begin{smallmatrix*}[r] U_{11}\end{smallmatrix*}\right]$}; \\
		};
		\draw (n1) -- ++(220:0.4cm) -- (a) ;
		\draw (n1) -- ++(260:0.4cm) -- (b) ;
		\draw (n1) -- ++(280:0.4cm) -- (c) ;
		\draw (n1) -- ++(320:0.4cm) -- (d)  ;
		
		\draw ($(n1)+(0,0.6cm)$) -- (n1);}
	
	\node (U) at ($(qmdd.west) + (-2,0.3)$) {$
		\begin{bNiceArray}{C|C}[create-medium-nodes, margin, columns-width = auto, code-after = {
		}, first-row, first-col, code-for-first-row = \color{gray}\scriptstyle , code-for-first-col = \color{gray}\scriptstyle]
		& \ket{0} & \ket{1} \\
		\ket{0} & U_{00} & U_{10}  \\\hline
		\ket{1} & U_{01} & U_{11}
		\end{bNiceArray}$};
	
	\node (Uname) at ($(U)+(-1.6,-0.1)$) {$U\colon$};
	\node[rotate=90, font=\scriptsize] (from) at ($(U)+(-1.25,-0.15)$) {\color{gray}{to}};
	\node[font=\scriptsize] (to) at ($(U)+(0.25,.6)$) {\color{gray}{from}};
	\node[font=\scriptsize, rotate=35] (to) at ($(U)+(-0.8,.5)$) {\color{gray}{$q_{n-1}$}};
	
	\draw[->, thick] ($(U.east) + (-0.1,-0.15)$) -- ++ (0.7,0);
	\end{tikzpicture}}
	\vspace*{-1mm}
	\caption{Decomposition scheme}
	\label{fig:first_level}
	\vspace*{-5mm}
\end{figure}

Consider a unitary matrix~\mbox{$U\in \mathbb{C}^{2^n\times 2^n}$}. Then, $U$ can be split into four $2^{n-1}\times 2^{n-1}$-sized sub-matrices $U_{ij}$ as shown on the left side of Fig.~\ref{fig:first_level}.
This splitting corresponds to the action of~$U$ depending on the value of the topmost qubit~$q_{n-1}$, i.e.,~ $U_{ij}$ describes how the rest of the system is transformed given that~$q_{n-1}$ is mapped from~$\ket{i}$ to~$\ket{j}$ for $i,j\in\{0,1\}$.
In the corresponding decision diagram, this manifests as a node with label $q_{n-1}$ and four successor nodes as shown on the right side of Fig.~\ref{fig:first_level}.

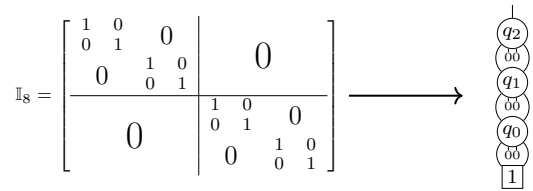
\begin{figure}[H]
	\centering
	\resizebox{0.8\linewidth}{!}{
	\begin{tikzpicture}
	\node (I) {\scalebox{0.9}{$\mathbb{I}_8= 
	\begin{bNiceArray}{CCCC|CCCC}[margin, nullify-dots]
	1 &0 & \Block{2-2}<\large>{0} & & \Block{4-4}<\LARGE>{0} & & & \\
	 0 & 1 & & & & &  \\
	\Block{2-2}<\large>{0} & & 1 &0 & & & \\
	 & & 0 & 1  &  & & \\\hline
	\Block{4-4}<\LARGE>{0} & & & & 1 &0 & \Block{2-2}<\large>{0} & \\
	& & & & 0& 1 \\
	& & &  &\Block{2-2}<\large>{0} &  & 1 & 0\\
	& &  & & &  & 0& 1
	\end{bNiceArray}$}};

	\node[draw, circle, inner sep=1pt] (q2) at ($(I.20) + (2.3,0)$) {$q_2$};
	\node[draw, circle, inner sep=1pt, below=0.25 of q2] (q1) {$q_1$};
	\node[draw, circle, inner sep=1pt, below=0.25 of q1] (q0) {$q_0$};
	\node[draw, rectangle, inner sep=2pt, below=0.25 of q0] (t) {$1$};
	\draw (q2) -- ++ (0, 0.4);
	
	\draw (q2.-135) to [out=-135,in=135] (q1.135);
	\draw (q2.-105) -- ++ (0, -0.05) node[below, inner sep = 0pt] {\tiny$0$};
	\draw (q2.-75) -- ++ (0, -0.05) node[below, inner sep = 0pt] {\tiny$0$};		
	\draw (q2.-45) to [out=-45,in=45] (q1.45);
	
	\draw (q1.-135) to [out=-135,in=135] (q0.135);
	\draw (q1.-105) -- ++ (0, -0.05) node[below, inner sep = 0pt] {\tiny$0$};
	\draw (q1.-75) -- ++ (0, -0.05) node[below, inner sep = 0pt] {\tiny$0$};		
	\draw (q1.-45) to [out=-45,in=45] (q0.45);
	
	\draw (q0.-135) to [out=-135,in=135] (t.135);
	\draw (q0.-105) -- ++ (0, -0.05) node[below, inner sep = 0pt] {\tiny$0$};
	\draw (q0.-75) -- ++ (0, -0.05) node[below, inner sep = 0pt] {\tiny$0$};
	\draw (q0.-45) to [out=-45,in=45] (t.45);
	\draw[->, thick] (I.0) -- ++ (1.6,0);
	\end{tikzpicture}}
	\caption{Identity matrix and DD for $n=3$}
	\label{fig:ident}
\end{figure}

This decomposition scheme can now be applied recursively until only single matrix entries (i.e., complex numbers) remain---eventually forming the decision diagram's terminal nodes.
During this process, identical sub-matrices can be represented by the same node which allows to capitalize on possible redundancies in the matrix $U$.
\begin{example}\label{ex:id}
The most compact decision diagram is achieved by the identity $\mathbb{I}$
since all sub-matrices $U_{00}$ and $U_{11}$ are recursively identical and \emph{all} sub-matrices $U_{01}$ and $U_{10}$ are solely composed of \mbox{0-entries}---yielding a decision diagram with exactly $n$ nodes as shown in Fig.~\ref{fig:ident} for three qubits\footnote{Sub-matrices only containing zero entries are represented as \mbox{$0$-stubs}.}. 
\end{example}

Further compaction is achieved by employing edge weights in order to unify sub-matrices only differing by a common factor.
As a consequence, one terminal node always suffices and, hence, is typically not counted towards the overall size (i.e.,~node count) of a decision diagram.

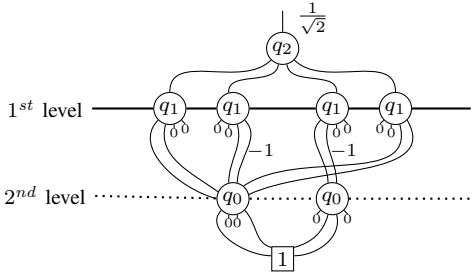
\begin{figure}[t]
	\centering
	\begin{tikzpicture}
	\node[draw, circle, inner sep=1pt] (q2) {$q_2$};
	\node[draw, circle, inner sep=1pt] at ($(q2) - (1.5,0.8)$) (q1a) {$q_1$};
	\node[draw, circle, inner sep=1pt] at ($(q2) - (0.66,0.8)$) (q1b) {$q_1$};
	\node[draw, circle, inner sep=1pt] at ($(q2) - (-0.66,0.8)$) (q1c) {$q_1$};
	\node[draw, circle, inner sep=1pt] at ($(q2) - (-1.5,0.8)$) (q1d) {$q_1$};
	\node[draw, circle, inner sep=1pt] at ($(q1b) - (0,1.2)$) (q0a) {$q_0$};
	\node[draw, circle, inner sep=1pt] at ($(q1c) - (0,1.2)$) (q0b) {$q_0$};
	\node[draw, rectangle, inner sep=2pt] at ($(q0a)!0.5!(q0b) - (0,0.8)$) (t) {$1$};
	
	\draw (q2) -- ++ (0, 0.5);
	\node (sqrt) at ($(q2) + (0.4, 0.4)$) {$\frac{1}{\sqrt{2}}$};
	
	\draw (q2.-135) to [out=-135,in=90] (q1a.90);
	\draw (q2.-105) to [out=-95,in=95] (q1b.105);
	\draw (q2.-75) to [out=-85,in=85] (q1c.75);
	\draw (q2.-45) to [out=-45,in=90] (q1d.90);
	
	\draw (q1a.-135) to [out=-135,in=165] (q0a.180);
	\draw (q1a.-105) to [out=-105,in=145] (q0a.155);
	\draw (q1a.-75) -- ++ (0, -0.05) node[below, inner sep = 0pt] {\tiny$0$};
	\draw (q1a.-45) -- ++ (0.05, -0.05) node[below, inner sep = 0pt] {\tiny$0$};
	
	\draw (q1b.-135) -- ++ (-0.05, -0.05) node[below, inner sep = 0pt] {\tiny$0$};
	\draw (q1b.-105) -- ++ (0, -0.05) node[below, inner sep = 0pt] {\tiny$0$};
	\draw (q1b.-75) to [out=-75,in=90] (q0a.105);
	\draw (q1b.-45) to [out=-45,in=90] (q0a.75);
	
	\draw (q1c.-135) to [out=-135,in=90] (q0b.105);
	\draw (q1c.-105) to [out=-105,in=90] (q0b.75);
	\draw (q1c.-75) -- ++ (0, -0.05) node[below, inner sep = 0pt] {\tiny$0$};
	\draw (q1c.-45) -- ++ (0.05, -0.05) node[below, inner sep = 0pt] {\tiny$0$};
	
	\draw (q1d.-135) -- ++ (-0.05, -0.05) node[below, inner sep = 0pt] {\tiny$0$};
	\draw (q1d.-105) -- ++ (0, -0.05) node[below, inner sep = 0pt] {\tiny$0$};
	\draw (q1d.-75) to [out=-75,in=35] (q0a.50);
	\draw (q1d.-45) to [out=-60,in=25] (q0a.15);
	
	\draw (q0a.-135) to [out=-135,in=180] (t.165);
	\draw (q0a.-105) -- ++ (0, -0.05) node[below, inner sep = 0pt] {\tiny$0$};
	\draw (q0a.-75) -- ++ (0, -0.05) node[below, inner sep = 0pt] {\tiny$0$};
	\draw (q0a.-45) to [out=-45,in=165] (t.135);
	
	\draw (q0b.-135) -- ++ (-0.05, -0.05) node[below, inner sep = 0pt] {\tiny$0$};
	\draw (q0b.-105) to [out=-105,in=15] (t.45);
	\draw (q0b.-75) to [out=-75,in=0] (t.15);
	\draw (q0b.-45) -- ++ (0.05, -0.05) node[below, inner sep = 0pt] {\tiny$0$};
	
	\node (m1a) at ($(q1b) + (0.37,-0.57)$) {\scriptsize$-1$};
	\node (m1b) at ($(q1c) + (0.15,-0.57)$) {\scriptsize$-1$};
	
	\node (l1) at ($(q1a)-(1.66,0)$) {$1^{\mathit{st}}$ level};
	\draw[thick] (l1) -- (q1a) -- (q1b) -- (q1c) -- (q1d) -- ++ (1.,0);
	\node (l2) at ($(l1)-(0,1.15)$) {$2^{\mathit{nd}}$ level};
	\draw[thick,dotted] (l2) -- (q0a) -- (q0b) -- ++ (1.88,0);
	\end{tikzpicture}
	\caption{Decision diagram for $U$}
	\label{fig:dd}
	\vspace*{-3mm}
\end{figure}

\begin{example}
	Consider again the matrix~$U$ as shown in Fig.~\ref{fig:u}. Recursively decomposing this matrix yields the decision diagram shown in Fig.~\ref{fig:dd}---representing the entire matrix~$U$.
	While, at the first decomposition level (indicated by a thick bold line in Fig.~\ref{fig:u}) all sub-matrices have a different structure, two nodes suffice to completely represent the entire functionality at the second decomposition level (indicated by a dotted line in Fig.~\ref{fig:u}).
	Due to the utilization of edge weights and exploitation of redundancies, just seven nodes suffice to capture the whole functionality of $U$.
\end{example}

Constructing decision diagrams in this fashion results in a \emph{canonic} representation of the functionality\footnote{Canonicity is achieved for a fixed variable order and under the assumption that a suitable normalization scheme is employed. For example, nodes can be normalized through division by the leftmost non-zero edge weight.}.
Moreover, vectors can be interpreted as matrices where each node's $U_{10}$ and $U_{11}$ successors are zero.

Decision diagrams do not only allow one to efficiently represent quantum functionality, but also to manipulate it. 
Here, we just illustrate how the most prominent operation---\mbox{matrix-matrix} multiplication---is conducted using decision diagrams.
However, many of the typical operations on matrices and vectors can directly be carried over for decision diagrams in a similar fashion. 
This includes, but is not limited to, addition, (conjugate) transposition, fidelity and (partial) trace calculation.
The multiplication of two matrices $U$ and $V$ is recursively broken down into sub-expressions according to
\begin{equation*}\resizebox{0.99\linewidth}{!}{$
\begin{bNiceMatrix}[]
U_{00} & U_{01} \\ U_{10} & U_{11}
\end{bNiceMatrix}\cdot
\begin{bNiceMatrix}[]
V_{00} & V_{01} \\ V_{10} & V_{11}
\end{bNiceMatrix} =
\begin{bNiceMatrix}[]
(U_{00}V_{00} + U_{01}V_{10}) & (U_{00}V_{01} + U_{01}V_{11}) \\ (U_{10}V_{00} + U_{11}V_{10}) & (U_{10}V_{01} + U_{11}V_{11})
\end{bNiceMatrix}$},
\end{equation*}
until only operations on complex numbers remain. Note that the individual sub-matrices exactly correspond to the respective successors of the currently considered decision diagram node. Eventually, this allows decision diagrams to efficiently represent and manipulate quantum functionality in many cases.

\vspace*{-3mm}
\section{Motivation}\label{sec:motivation}
As reviewed in Section~\ref{sec:intro}, the typical design flow for realizing a quantum algorithm on a real quantum computer requires several steps in which the desired functionality is decomposed, mapped, and optimized. During all these steps, it has to be ensured that the functionality of the correspondingly resulting circuit descriptions does not change---motivating equivalence checking for quantum circuits.
In this section, we first give an explicit description of this problem. %
Then, we review how equivalence checking is conducted to date and discuss why current solutions are still unsatisfactory.
Those discussions provide the overall motivation of our work.

\subsection{The Quantum Circuit Equivalence Checking Problem}\label{sec:considered}
The functionality of a quantum algorithm is most prominently described in the form of a quantum circuit---potentially consisting of high-level operations, such as Grover iterations~\cite{groverFastQuantumMechanical1996}, the quantum Fourier transform~\cite{nielsenQuantumComputationQuantum2010}, etc.
Technically, one could also consider other \enquote{functional descriptions} such as a unitary matrix, a decision diagram, or a tensor network.
However, those are usually internal representations used by tools and hardly a functional description provided by a designer.
Consequently, equivalence checking in the domain of quantum computing---as we consider it in this work---is about proving that two quantum circuits~$G$ and~$G^\prime$ are functionally equivalent (i.e., realize the same function), or to show the non-equivalence of these circuits by means of a counterexample. 

To this end, consider two quantum circuits \mbox{$G=g_0\dots g_{m-1}$} and \mbox{$G^\prime=g^\prime_0\dots g^\prime_{m^\prime-1}$} operating on $n$ qubits. 
Then, as reviewed in the previous section, the functionality of each circuit can uniquely be described %
by the respective system matrices \mbox{$U=U_{m-1}\cdots U_0$} and \mbox{$U^\prime=U_{m^\prime-1}^\prime\cdots U^\prime_0$}, where the matrices~$U^{(\prime)}_i$ describe the functionality of the \mbox{$i$-th}~gate of the respective circuit (with $ 0 \le i < m^{(\prime)}$). %
Consequently, deciding the equivalence of both computations amounts to comparing the system matrices $U$ and $U^\prime$. More precisely, $U$ and $U^\prime$ are considered equivalent, if they at most differ by a global phase factor (which is fundamentally unobservable~\cite{nielsenQuantumComputationQuantum2010}), i.e.,~\mbox{$U = e^{i\alpha}U^\prime$} with $\alpha\in[0,2\pi)$.

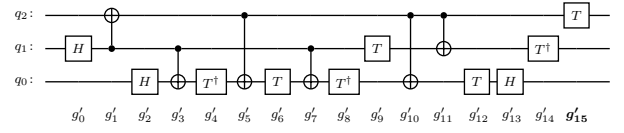
\begin{figure}[tb]
	\centering
	\resizebox{0.92\linewidth}{!}{
		\begin{quantikz}[column sep={0.7cm,between origins}, row sep={0.7cm,between origins}, ampersand replacement=\&]
			\lstick{$q_2 \colon$} \& \qw \& \targ{} \& \qw \& \qw  \& \qw \& \ctrl{2} \& \qw \& \qw \& \qw \& \qw \& \ctrl{2} \& \ctrl{1} \& \qw \& \qw \& \qw \& \gate{T} \& \qw \\
			\lstick{$q_1\colon$}  \& \gate{H} \& \ctrl{-1}\& \qw \& \ctrl{1} \&\qw \& \qw \& \qw \& \ctrl{1} \& \qw \& \gate{T} \& \qw \& \targ{} \& \qw \& \qw \& \gate{T^\dag} \& \qw \& \qw \\
			\lstick{$q_0\colon$} \& \qw  \& \qw \& \gate{H} \& \targ{} \& \gate{T^\dag} \& \targ{} \& \gate{T} \& \targ{} \& \gate{T^\dag} \& \qw \& \targ{} \& \qw \& \gate{T} \& \gate{H} \& \qw \& \qw \& \qw \\
			\& \lstick[label style={xshift=0.3cm}]{$g_0^\prime$} \& \lstick[label style={xshift=0.3cm}]{$g_1^\prime$} \& \lstick[label style={xshift=0.3cm}]{$g_2^\prime$} \& \lstick[label style={xshift=0.3cm}]{$g_3^\prime$} \& \lstick[label style={xshift=0.3cm}]{$g_4^\prime$}\& \lstick[label style={xshift=0.3cm}]{$g_5^\prime$}\& \lstick[label style={xshift=0.3cm}]{$g_6^\prime$}\& \lstick[label style={xshift=0.3cm}]{$g_7^\prime$}\& \lstick[label style={xshift=0.3cm}]{$g_8^\prime$}\& \lstick[label style={xshift=0.3cm}]{$g_9^\prime$}\& \lstick[label style={xshift=0.35cm}]{$g_{10}^\prime$}\& \lstick[label style={xshift=0.35cm}]{$g_{11}^\prime$}\& \lstick[label style={xshift=0.4cm}]{$g_{12}^\prime$} \& \lstick[label style={xshift=0.4cm}]{$g_{13}^\prime$} \& \lstick[label style={xshift=0.4cm}]{$g_{14}^\prime$} \& \lstick[label style={xshift=0.4cm}]{$\bm{g_{15}^\prime}$} 
	\end{quantikz}}
	\caption{Alternative circuit realization $G^\prime$}
	\label{fig:Gprime}
\end{figure}

\begin{example}\label{ex:eq}
	Consider again the circuit $G$ from Fig.~\ref{fig:g} and an alternative realization $G^\prime$ as shown in Fig.~\ref{fig:Gprime}. 
	Since the functionality of %
	$G^\prime$ exhibits the same system matrix $U$ as shown before in Fig.~\ref{fig:u}, %
	both circuits $G$ and $G^\prime$ are indeed equivalent.
\end{example}

If $U$ and $U^\prime$ differ in any column~$i$ (by more than a global phase factor~$e^{i\alpha}$), then the corresponding circuits $G$ and $G^\prime$ are not equivalent and~\ket{i} serves as a counterexample showing that %
\[U \ket{i} = \ket{u_i} \; \not\equiv\; \ket{u_i^\prime} = U^\prime \ket{i}.\]
Here, the fidelity $\mathcal{F}$ between two states \ket{x} and \ket{y} is typically used as a similarity measure for comparing quantum states, where $\mathcal{F}$ is calculated as the squared overlap between the states, i.e., $\mathcal{F} = \vert\braket{x}{y}\vert^2\in[0,1]$. Two states are considered equivalent if the fidelity between them is $1$ (up to a given tolerance $\varepsilon$).

\begin{example}\label{ex:noneq}
	Consider the same scenario as in Ex.~\ref{ex:eq}, but additionally assume that, due to an %
	error, the last $\qop{T}$-gate of~$G^\prime$ (i.e.,~$g_{15}^\prime$) is not applied (yielding a %
	new circuit $\tilde{G}^\prime$). Then, a new functionality results, which is described by the %
	system matrix $\tilde{U}^\prime$ %
	shown in Fig.~\ref{fig:Uprime}. 
	Since $U$ and $\tilde{U}^\prime$ are obviously not identical anymore, the circuits~$G$ and~$\tilde{G}^\prime$ have been shown to be non-equivalent.
	Moreover, since both matrices differ, e.g., in column four, $\ket{4}$ serves as a counterexample showing that 
	\[U \ket{4} = \ket{u_4} = \tfrac{1}{\sqrt{2}} [0\;0\;1\;0\;0\;0\;1\;0]^\top,\] 
	while 
	\[U^\prime \ket{4} = \ket{u_4^\prime} = \tfrac{1}{\sqrt{2}} [0\;0\;1\;0\;0\;0\;\tfrac{1-i}{\sqrt{2}}\;0]^\top.\]
	It holds that $\mathcal{F}(U \ket{4},\, U^\prime \ket{4}) = \mathcal{F}(\ket{u_4},\ket{u_4^\prime}) \approx 0.92 < 1$.
\end{example}

\begin{figure}[tb]
	\centering
	\resizebox{0.78\linewidth}{!}{
		$\frac{1}{\sqrt{2}}\begin{bNiceArray}{RR:RR|RR:RR}[create-medium-nodes, margin, columns-width = auto]
		1 & 0 & 1 & 0 & \bm{0} & 0 & 0 & 0 \\
		0 & 1 & 0 & 1 & \bm{0} & 0 & 0 & 0 \\\hdottedline
		0 & 0 & 0 & 0 & \bm{1} & 0 & -1 & 0 \\
		0 & 0 & 0 & 0 & \bm{0} & 1 & 0 & -1\\\hline
		0 & \color{red}{-\omega^3} & 0 & \color{red}{\omega^3} & \bm{0} & 0 & 0 & 0 \\
		\color{red}{-\omega^3} & 0 & \color{red}{\omega^3} & 0 & \bm{0} & 0 & 0 & 0 \\\hdottedline
		0 & 0 & 0 & 0 & \color{red}{\bm{-\omega^3}} & 0 & \color{red}{-\omega^3} & 0 \\
		0 & 0 & 0 & 0 & \bm{0} & \color{red}{-\omega^3} & 0 & \color{red}{-\omega^3}\\
		\end{bNiceArray}$
}
	\caption{Erroneous circuit matrix $\tilde{U}^\prime$ ($\omega=\frac{1+i}{\sqrt{2}}$)}
	\label{fig:Uprime}
\end{figure}

Unfortunately, the whole functionality $U$ (and similarly~$U^\prime$) is not readily available for performing this comparison, but has to be constructed from the individual gate descriptions~$g_i$---requiring the subsequent  matrix-matrix multiplications 
\[
U^{(0)} = U_0,\quad U^{(j)} = U_{j} \cdot U^{(j-1)} \mbox{ for } j=1,\dots,m-1
\]
to construct the whole system matrix $U = U^{(m-1)}$.
While conceptually simple, this quickly constitutes an extremely complex task due to the exponential size of the involved matrices  (and vectors) with respect to the number of qubits.
 In fact, equivalence checking of quantum circuits has been shown to be QMA-complete~\cite{janzingNonidentityCheckQMAcomplete2005}.

\subsection{Related Work}\label{sec:related}

In order to tackle the underlying complexity, equivalence checking of quantum circuits has been intensively considered in the past. 
Inspired by methods for equivalence checking of classical circuits~\cite{brandVerificationLargeSynthesized1993, molitorEquivalenceCheckingDigital2010, marques-silvaCombinationalEquivalenceChecking1999, jhaEquivalenceCheckingUsing1997}, methods based on re-writing~\cite{yamashitaFastEquivalencecheckingQuantum2010, duncanGraphtheoreticSimplificationQuantum2019}, Boolean satisfiability~\cite{willeCompactEfficientSAT2013}, and decision diagrams~\mbox{\cite{wangXQDDbasedVerificationMethod2008,niemannQMDDsEfficientQuantum2016,viamontesCheckingEquivalenceQuantum2007,niemannEquivalenceCheckingMultilevel2014}}
have been proposed.
However, approaches based on re-writing suffer from the fact that they are either incomplete, provide inconclusive results in case of non-equivalence, or may require the enumeration of an exponential number of possibilities to check. 
Solvers for Boolean satisfiability may cope with these problems but, in the case of quantum circuits, face the problem that an infinite number of quantum states needs to be encoded.\footnote{This problem has partially been addressed in~\cite{willeCompactEfficientSAT2013} by employing a detailed structural analysis that restricts the number of possible states to a finite one. However, the resulting encoding does not support key features of quantum circuits such as entanglement and, hence, is not applicable for almost all relevant quantum circuits.}
Because of that, approaches based on decision diagrams (also called \emph{DD-based equivalence checking}) still constitute the state of the art for equivalence checking of quantum circuits.

Indeed, decision diagrams offer several benefits for this task: As reviewed in Section~\ref{sec:dds}, they provide means for compact representation and efficient manipulation of the respective functionality realized by a quantum circuit. Moreover, most decision diagrams 
provide a canonical way of representing quantum functionality (usually with respect to a given variable order and normalization scheme). Because of that, once the functionalities of different quantum circuits are represented as decision diagrams, checking their equivalence can be conducted in constant time by simply comparing their root edges (and the associated weights in case both circuits differ only by a global phase factor). 

\begin{example}\label{ex:dd_based_ec}
	Consider again the two quantum circuits~$G$~and~$G^\prime$ from Ex.~\ref{ex:eq}. 
	Creating decision diagrams for both circuits yields the representation as shown in Fig.~\ref{fig:dd}.
	Since representations for both, $U$ and $U^\prime$, would point to the (same) root node of this decision diagram,
	$G$~and~$G^\prime$ are proven to be functionally equivalent. 
\end{example}

However, despite the benefits discussed above, \mbox{DD-based} equivalence checking of quantum circuits as conducted thus far still has significant shortcomings. In fact, representing the entire functionality of a quantum circuit still might be exponential in the worst case. Even if the representation of the overall functionality of a circuit might be compact, intermediate results may require significantly more space. As an example, the final decision diagram discussed in Ex.~\ref{ex:dd_based_ec} and shown in Fig.~\ref{fig:dd} is composed of seven nodes; however, intermediate decision diagrams generated during the construction required up to nine nodes.\footnote{An increase by two nodes might not seem substantial. However, for realistic quantum circuits which represent much more complex functionality than this example, the difference in the number of nodes between the final and intermediate decision diagrams easily sums up to several orders of magnitudes.} 
Moreover, the generation of a counterexample (which is supposed to be provided in case two quantum circuits are not equivalent) requires further potentially expensive manipulations of the decision diagrams. In fact, in order to generate such a counterexample, the ``difference'' between the two non-equivalent decision diagrams has to be determined. To this end, one decision diagram has to be inverted (i.e., the conjugate-transpose representation has to be generated) and multiplied with the other decision diagram---requiring \mbox{non-trivial} operations on potentially large representations. Then, any path in the resulting difference decision diagram containing a node not resembling the identity constitutes a counterexample.

\begin{example}\label{ex:cex}
	Consider again the quantum circuit $G$ shown in Fig.~\ref{fig:g} and the erroneous circuit $\tilde{G^\prime}$ from Ex.~\ref{ex:noneq}.
	Then, the decision diagram representation of $\tilde{U^\prime}$ would not point to the same root node as the decision diagram for $U$---from which it can be concluded that they are not equivalent. 
	In order to generate counterexamples, the difference of both circuits has to be determined---requiring the inversion of the decision diagram for $\tilde{U}^\prime$ (i.e., calculating the conjugate-transposed) and multiplication with the decision diagram for~$U$ (see~Fig.~\ref{fig:dd}).
	Eventually, this results in the decision diagram shown in Fig.~\ref{fig:diff}.
	From this, one may obtain, e.g., the counterexample $\ket{(100)_2} = \ket{4}$ as observed previously in Ex.~\ref{ex:noneq}, since the corresponding path contains the node $q_2$---which does not resemble the identity due to the rightmost edge weight being equal to $\nicefrac{1}{\sqrt{2}}(1+i)\neq 1$.
\end{example}

\begin{figure}[t]
	\centering
	\begin{tikzpicture}
	\node[draw, circle, inner sep=1pt] (q2) {$q_2$};
	\node[draw, circle, inner sep=1pt, below=0.25 of q2] (q1) {$q_1$};
	\node[draw, circle, inner sep=1pt, below=0.25 of q1] (q0) {$q_0$};
	\node[draw, rectangle, inner sep=2pt, below=0.25 of q0] (t) {$1$};
	\draw (q2) -- ++ (0, 0.4);
	
	\draw (q2.-135) to [out=-135,in=135] (q1.135);
	\draw (q2.-105) -- ++ (0, -0.05) node[below, inner sep = 0pt] {\tiny$0$};
	\draw (q2.-75) -- ++ (0, -0.05) node[below, inner sep = 0pt] {\tiny$0$};		
	\draw[red,thick] (q2.-45) to [out=-45,in=45] (q1.45);
	
	\draw (q1.-135) to [out=-135,in=135] (q0.135);
	\draw (q1.-105) -- ++ (0, -0.05) node[below, inner sep = 0pt] {\tiny$0$};
	\draw (q1.-75) -- ++ (0, -0.05) node[below, inner sep = 0pt] {\tiny$0$};		
	\draw (q1.-45) to [out=-45,in=45] (q0.45);
	
	\draw (q0.-135) to [out=-135,in=135] (t.135);
	\draw (q0.-105) -- ++ (0, -0.05) node[below, inner sep = 0pt] {\tiny$0$};
	\draw (q0.-75) -- ++ (0, -0.05) node[below, inner sep = 0pt] {\tiny$0$};
	\draw (q0.-45) to [out=-45,in=45] (t.45);
	\node (om) at ($(q2)!0.5!(q1) + (0.4,0)$) {\color{red}{$\omega$}};
	\end{tikzpicture}
	\vspace*{-1mm}
	\caption{Difference DD of $G$ and $\tilde{G^\prime}$}
	\label{fig:diff}
	\vspace*{-5mm}
\end{figure}
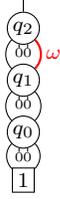

\begin{figure*}[t]
	\centering
	\def\scale{0.08}
\begin{tikzpicture}
\matrix[ampersand replacement=\&, nodes={inner sep=0pt, anchor=west}, row sep=0.1cm, font=\small, column sep=0.75em] (dds1) {
	\node[] (descr) {Gate application sequence: }; \&
	\node[] (g0) {$\mathbf{\mathbb{I}}$}; \&
	\node[] (g1) {$g_0$}; \&
	\node[] (g2) {$g_{0}^{\prime \dag}$}; \&
	\node[] (g3) {$g_1$}; \&
	\node[] (g4) {$g_{1}^{\prime \dag}$}; \&
	\node[] (g5) {$g_{2}^{\prime \dag}$}; \&
	\node[] (g6) {$g_{3}^{\prime \dag}$}; \&
	\node[] (g7) {$g_{4}^{\prime \dag}$}; \&
	\node[] (g8) {$g_2$};\&
	\node[] (g9) {$g_3$};\&
	\node[] (g10) {$g_{5}^{\prime \dag}$}; \&
	\node[] (g11) {$g_{6}^{\prime \dag}$}; \&
	\node[] (g12) {$g_{7}^{\prime \dag}$}; \&
	\node[] (g13) {$g_{8}^{\prime \dag}$}; \&
	\node[] (g14) {$g_{9}^{\prime \dag}$}; \&
	\node[] (g15) {$g_{10}^{\prime \dag}$}; \&
	\node[] (g16) {$g_{11}^{\prime \dag}$}; \&
	\node[] (g17) {$g_{12}^{\prime \dag}$}; \&
	\node[] (g18) {$g_{13}^{\prime \dag}$}; \&
	\node[] (g19) {$g_{14}^{\prime \dag}$}; \&
	\node[] (g20) {$g_{15}^{\prime \dag}$}; \\
	\\
	\node[] (count) {DD node count: }; \&
	\node[] (n0) {$\mathbf{3}$}; \&
	\node[] (n1) {$\mathbf{3}$}; \&
	\node[] (n2) {$\mathbf{3}$}; \&
	\node[] (n3) {$\mathbf{4}$}; \&
	\node[] (n4) {$\mathbf{3}$}; \&
	\node[] (n5) {$\mathbf{3}$}; \&
	\node[] (n6) {$\mathbf{4}$};\&
	\node[] (n7) {$\mathbf{4}$};\&
	\node[] (n8) {$\mathbf{6}$};\&
	\node[] (n9) {$\mathbf{6}$};\&
	\node[] (n10) {$\mathbf{7}$};\&
	\node[] (n11) {$\mathbf{7}$};\&
	\node[] (n12) {$\mathbf{7}$};\&
	\node[] (n13) {$\mathbf{5}$};\&
	\node[] (n14) {$\mathbf{5}$};\&
	\node[] (n15) {$\mathbf{4}$};\&
	\node[] (n16) {$\mathbf{3}$};\&
	\node[] (n17) {$\mathbf{3}$};\&
	\node[] (n18) {$\mathbf{3}$};\&
	\node[] (n19) {$\mathbf{3}$};\&
	\node[] (n20) {$\mathbf{3}$};\\
};
\draw[thick] (g0) -- (g1) -- (g2) -- (g3) -- (g4) -- (g5) -- (g6) -- (g7) -- (g8) -- (g9) -- (g10) -- (g11) -- (g12) -- (g13) -- (g14) -- (g15) -- (g16) -- (g17) -- (g18) -- (g19) -- (g20);
\draw[thick] (n0) -- (n1) -- (n2) -- (n3) -- (n4) -- (n5) -- (n6) -- (n7) -- (n8) -- (n9) -- (n10) -- (n11) -- (n12) -- (n13) -- (n14) -- (n15) -- (n16) -- (n17) -- (n18) -- (n19) -- (n20);
\end{tikzpicture}
\caption{Illustration of the general idea: $G \shortrightarrow \mathbb{I} \shortleftarrow G^{\prime}$ with $G$ and $G^{\prime}$ from Ex.~\ref{ex:dd_based_ec} using decision diagrams}
\label{fig:g_ginverse_id}
\vspace*{-5mm}
\end{figure*}
\vspace*{-3mm}
\section{General Ideas}\label{sec:general}
\vspace*{-1mm}
In this section, we illustrate the general ideas developed in this work to address the shortcomings of the related work discussed above and, indeed, to solve the equivalence checking problem dramatically faster than ever before---in many cases even just with a single simulation run. 
To this end, we utilize %
certain characteristics of quantum computing. 
While those,
at a first glance, make the problem of verification harder
compared to the classical realm (circuits have to be supported which do not only rely on $0$'s and $1$'s, but also on superposition or entanglement), they also offer potential which has not really been exploited yet. 
In the following, %
we sketch %
this unearthed potential before dedicated equivalence checking schemes and flows resulting from these ideas are covered in the remaining sections of this paper. 
\vspace*{-2mm}
\subsection{Exploiting Reversibility}\label{sec:gen_aspdac}
Many classical logic operations are irreversible (e.g.,~\mbox{$x \wedge y = 0$} does not allow one to determine the precise values of $x$ and $y$).
As there is no bijective mapping between input and output states, in general, the concept of the \emph{inverse} of a classical operation (or a sequence thereof) does not make sense.
In contrast, all quantum operations are inherently \emph{reversible}. 
Consider an operation $g$ described by the unitary matrix $U$.
Then, its inverse $U^{-1}$ is efficiently calculated as the conjugate-transpose $U^\dag$.
Given a sequence of $m$ operations $g_0, \dots, g_{m-1}$ with associated matrices $U_0,\dots, U_{m-1}$,
the inverse of the corresponding system matrix $U=U_{m-1}\cdots U_0$ is derived by reversing the operations' order and inverting each individual operation, i.e., 
$U^{-1} =U^\dag= U_{0}^\dag \cdots U_{m-1}^\dag$.

Now consider two quantum circuits $G$ and $G^\prime$. In case both circuits are functionally equivalent, this allows for the conclusion that concatenating one circuit with the inverse of the other realizes the identity function $\mathbb{I}$, i.e.,
\[
G \cdot G^{\prime -1} = {G^{\prime -1} \cdot G} = G^{-1} \cdot G^\prime = G^\prime \cdot G^{-1} \equiv \mathbb{I}.\footnote{Throughout this paper we particularly focus on the arrangement $G^{\prime -1} \cdot G$ as it results in the ``intuitive'' order of matrix multiplication $U\cdot U^{\prime\dag}$. However, all findings in this paper hold for any possible arrangement of $G$ and $G^\prime$.}
\]
Since, as discussed in Section~\ref{sec:dds}, the identity constitutes the best case for decision diagrams (the identity can be represented by a decision diagram of linear size), %
this offers significant potential.

Unfortunately, creating such a concatenation in a naive fashion, e.g., by computing $U \cdot U^{\prime\dag}$ hardly yields any benefits compared to the state of the art. 
That is, because, even if the final decision diagram would be as compact as possible, the full (and potentially exponential) decision diagrams representing $U\equiv G$ and $U^{\prime \dag}\equiv G^{\prime -1}$ would still be generated as intermediate results.
Even if the whole sequence of $m+m^\prime$ operations is considered as one huge quantum circuit whose functionality is to be constructed and compared to the identity, this would still entail the construction of the full decision diagram of one circuit for the first $m^{(\prime)}$ operations.
\begin{example}\label{ex:naivecombination}
	Consider again the two circuits $G$ and $G^{\prime}$ from Fig.~\ref{fig:g} and Fig.~\ref{fig:Gprime} as discussed before in Ex.~\ref{ex:dd_based_ec}. Concatenating $G$ and $G^{\prime -1}$ yields a quantum circuit $\tilde{G} = G \cdot G^{\prime -1}$ with \mbox{$4+16=20$} gates. Since both computations are equivalent (see~Ex.~\ref{ex:eq}) constructing the functionality of $\tilde{G}$ would eventually yield an identity decision diagram as shown in Fig.~\ref{fig:ident}. However, during the construction, the first four operations would essentially construct the decision diagram for $G$ as shown in Fig.~\ref{fig:dd}. Thus, there is no real benefit over the state of the art when conducting equivalence checking in this fashion.
\end{example}

Instead, the full potential of this observation is utilized if the associativity of the respective multiplications is fully exploited.
More precisely, given two quantum circuits $G$ and $G^\prime$, it holds that
\vspace*{-3mm}\begin{align*}
G^{\prime -1} \cdot G &=  {(g^{\prime -1}_{m^\prime-1}\dots g^{\prime -1}_0) \cdot (g_0\dots g_{m-1})} \\
&\equiv {(U_{m-1}\cdots U_0)\cdot(U_0^{\prime \dag} \cdots U_{m^\prime -1}^{\prime \dag})} \\
& = {U_{m-1}\cdots U_0\cdot \mathbb{I} \cdot U_0^{\prime \dag} \cdots U_{m^\prime-1}^{\prime \dag}} \\
&\eqqcolon G \shortrightarrow \mathbb{I} \shortleftarrow G^{\prime}.\vspace*{-3mm}
\end{align*}
Here, $G \shortrightarrow \mathbb{I} \shortleftarrow G^{\prime}$ symbolizes that, starting from the identity~$\mathbb{I}$, either gates from $G$ can be ``applied from the left'', or (inverted) gates from $G^{\prime}$ can be ``applied from the right''.
If the respective gates of $G$ and $G^{\prime}$ are applied in a fashion frequently yielding the identity, the entire equivalence checking process can be conducted with rather small (intermediate) decision diagrams only. This is illustrated by the following example. 

\begin{example} \label{ex:proposed_approach}
	Consider again the two circuits $G$ and $G^{\prime}$ from Ex.~\ref{ex:naivecombination} and assume that, starting with a decision diagram representing the identity, gates from~$G$ and~$G^{\prime}$ are 
	applied as sketched in Fig.~\ref{fig:g_ginverse_id}. Here, the top of the figure indicates the respectively applied gates, 
	and the bottom provides the corresponding sizes of the decision diagrams. 
	As can be seen, applying the gates from~$G$ and~$G^\prime$ in a particular order ``from the left'' and ``from the right'', respectively,
	frequently yields situations where the impact of a gate from circuit~$G$  (potentially increasing the size of the decision diagram) is reverted by multiplications with inverted gates from $G^{\prime}$ (potentially decreasing the size of the decision diagram back to the representation of the identity function). 
\end{example}

Moreover, even if the considered circuits~$G$ and~$G^\prime$ are \emph{not} functionally equivalent (and, hence, identity is not achieved), the observations from above still promise improvements compared to creating the complete decision diagrams for~$G$ and~$G^\prime$. This is, because in this case, %
the result of $G\shortrightarrow \mathbb{I}\shortleftarrow G^{\prime}$ inherently provides an efficient representation of the circuit's difference that allows one to obtain %
counterexamples almost ``for free'' (while \mbox{state-of-the-art} solutions have to explicitly create those using additional inversion and multiplication operations as discussed in Section~\ref{sec:related}).  

\begin{example}
	Consider again the scenario of Ex.~\ref{ex:cex}.
	In contrast to the state-of-the-art equivalence checking routine, applying the general idea proposed above follows the same steps as illustrated in Fig.~\ref{fig:g_ginverse_id} up to the very last multiplication---which is dropped.
	This precisely leads to the decision diagram obtained in Ex.~\ref{ex:cex} and shown in Fig.~\ref{fig:diff}.
	Thus, counterexamples can easily be derived from this decision diagram (see~Ex.~\ref{ex:cex}).
\end{example}

Overall, exploiting this characteristic and following those ideas, equivalence checking of two quantum circuits can be conducted much more efficiently and compactly than before. %
But determining when to apply gates from~$G$ and when to apply (inverted) gates from~$G^{\prime}$ is not at all obvious. 
Corresponding strategies for this purpose will be presented and evaluated in Section~\ref{sec:strategies} and Section~\ref{sec:experiments}, respectively.

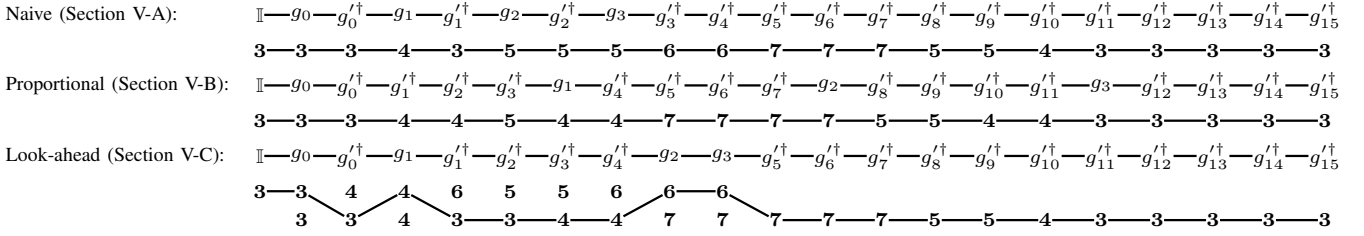
\begin{figure*}[th]
	\centering
	\resizebox{\linewidth}{!}{
	\begin{tikzpicture}
	\matrix[ampersand replacement=\&, nodes={inner sep=0pt, anchor=center}, row sep=0.2cm, column sep=0.3cm, font=\scriptsize] (strat) {
		\node[anchor=west] (naive) {Naive (Section~\ref{sec:naive}): }; \& 
		\node[] (g0) {$\mathbf{\mathbb{I}}$}; \&
		\node[] (g1) {$g_0$}; \&
		\node[] (g2) {$g_{0}^{\prime\dag}$}; \&
		\node[] (g3) {$g_1$}; \&
		\node[] (g4) {$g_{1}^{\prime\dag}$}; \&
		\node[] (g5) {$g_2$}; \&
		\node[] (g6) {$g_{2}^{\prime\dag}$}; \&
		\node[] (g7) {$g_3$}; \&
		\node[] (g8) {$g_{3}^{\prime\dag}$};\&
		\node[] (g9) {$g_{4}^{\prime\dag}$};\&
		\node[] (g10) {$g_{5}^{\prime\dag}$}; \&
		\node[] (g11) {$g_{6}^{\prime\dag}$}; \&
		\node[] (g12) {$g_{7}^{\prime\dag}$}; \&
		\node[] (g13) {$g_{8}^{\prime\dag}$}; \&
		\node[] (g14) {$g_{9}^{\prime\dag}$}; \&
		\node[] (g15) {$g_{10}^{\prime\dag}$}; \&
		\node[] (g16) {$g_{11}^{\prime\dag}$}; \&
		\node[] (g17) {$g_{12}^{\prime\dag}$}; \&
		\node[] (g18) {$g_{13}^{\prime\dag}$}; \&
		\node[] (g19) {$g_{14}^{\prime\dag}$}; \&
		\node[] (g20) {$g_{15}^{\prime\dag}$}; \\ \&
		\node[] (n0) {$\mathbf{3}$}; \&
		\node[] (n1) {$\mathbf{3}$}; \&
		\node[] (n2) {$\mathbf{3}$}; \&
		\node[] (n3) {$\mathbf{4}$}; \&
		\node[] (n4) {$\mathbf{3}$}; \&
		\node[] (n5) {$\mathbf{5}$}; \&
		\node[] (n6) {$\mathbf{5}$};\&
		\node[] (n7) {$\mathbf{5}$};\&
		\node[] (n8) {$\mathbf{6}$};\&
		\node[] (n9) {$\mathbf{6}$};\&
		\node[] (n10) {$\mathbf{7}$};\&
		\node[] (n11) {$\mathbf{7}$};\&
		\node[] (n12) {$\mathbf{7}$};\&
		\node[] (n13) {$\mathbf{5}$};\&
		\node[] (n14) {$\mathbf{5}$};\&
		\node[] (n15) {$\mathbf{4}$};\&
		\node[] (n16) {$\mathbf{3}$};\&
		\node[] (n17) {$\mathbf{3}$};\&
		\node[] (n18) {$\mathbf{3}$};\&
		\node[] (n19) {$\mathbf{3}$};\&
		\node[] (n20) {$\mathbf{3}$};\\
		\node[anchor=west] (prop) {Proportional (Section~\ref{sec:proportional}): }; \& 
		\node[] (g0p) {$\mathbf{\mathbb{I}}$}; \&
		\node[] (g1p) {$g_0$}; \&
		\node[] (g2p) {$g_{0}^{\prime\dag}$}; \&
		\node[] (g3p) {$g_{1}^{\prime\dag}$}; \&
		\node[] (g4p) {$g_{2}^{\prime\dag}$}; \&
		\node[] (g5p) {$g_{3}^{\prime\dag}$}; \&
		\node[] (g6p) {$g_1$}; \&
		\node[] (g7p) {$g_{4}^{\prime\dag}$}; \&
		\node[] (g8p) {$g_{5}^{\prime\dag}$};\&
		\node[] (g9p) {$g_{6}^{\prime\dag}$};\&
		\node[] (g10p) {$g_{7}^{\prime\dag}$}; \&
		\node[] (g11p) {$g_2$}; \&
		\node[] (g12p) {$g_{8}^{\prime\dag}$}; \&
		\node[] (g13p) {$g_{9}^{\prime\dag}$}; \&
		\node[] (g14p) {$g_{10}^{\prime\dag}$}; \&
		\node[] (g15p) {$g_{11}^{\prime\dag}$}; \&
		\node[] (g16p) {$g_3$}; \&
		\node[] (g17p) {$g_{12}^{\prime\dag}$}; \&
		\node[] (g18p) {$g_{13}^{\prime\dag}$}; \&
		\node[] (g19p) {$g_{14}^{\prime\dag}$}; \&
		\node[] (g20p) {$g_{15}^{\prime\dag}$}; \\ \&
		\node[] (n0p) {$\mathbf{3}$}; \&
		\node[] (n1p) {$\mathbf{3}$}; \&
		\node[] (n2p) {$\mathbf{3}$}; \&
		\node[] (n3p) {$\mathbf{4}$}; \&
		\node[] (n4p) {$\mathbf{4}$}; \&
		\node[] (n5p) {$\mathbf{5}$}; \&
		\node[] (n6p) {$\mathbf{4}$};\&
		\node[] (n7p) {$\mathbf{4}$};\&
		\node[] (n8p) {$\mathbf{7}$};\&
		\node[] (n9p) {$\mathbf{7}$};\&
		\node[] (n10p) {$\mathbf{7}$};\&
		\node[] (n11p) {$\mathbf{7}$};\&
		\node[] (n12p) {$\mathbf{5}$};\&
		\node[] (n13p) {$\mathbf{5}$};\&
		\node[] (n14p) {$\mathbf{4}$};\&
		\node[] (n15p) {$\mathbf{4}$};\&
		\node[] (n16p) {$\mathbf{3}$};\&
		\node[] (n17p) {$\mathbf{3}$};\&
		\node[] (n18p) {$\mathbf{3}$};\&
		\node[] (n19p) {$\mathbf{3}$};\&
		\node[] (n20p) {$\mathbf{3}$};\\
		\node[anchor=west] (look) {Look-ahead (Section~\ref{sec:lookahead}): }; \& 
		\node[] (g0l) {$\mathbf{\mathbb{I}}$}; \&
		\node[] (g1l) {$g_0$}; \&
		\node[] (g2l) {$g_{0}^{\prime\dag}$}; \&
		\node[] (g3l) {$g_1$}; \&
		\node[] (g4l) {$g_{1}^{\prime\dag}$}; \&
		\node[] (g5l) {$g_{2}^{\prime\dag}$}; \&
		\node[] (g6l) {$g_{3}^{\prime\dag}$}; \&
		\node[] (g7l) {$g_{4}^{\prime\dag}$}; \&
		\node[] (g8l) {$g_2$};\&
		\node[] (g9l) {$g_3$};\&
		\node[] (g10l) {$g_{5}^{\prime\dag}$}; \&
		\node[] (g11l) {$g_{6}^{\prime\dag}$}; \&
		\node[] (g12l) {$g_{7}^{\prime\dag}$}; \&
		\node[] (g13l) {$g_{8}^{\prime\dag}$}; \&
		\node[] (g14l) {$g_{9}^{\prime\dag}$}; \&
		\node[] (g15l) {$g_{10}^{\prime\dag}$}; \&
		\node[] (g16l) {$g_{11}^{\prime\dag}$}; \&
		\node[] (g17l) {$g_{12}^{\prime\dag}$}; \&
		\node[] (g18l) {$g_{13}^{\prime\dag}$}; \&
		\node[] (g19l) {$g_{14}^{\prime\dag}$}; \&
		\node[] (g20l) {$g_{15}^{\prime\dag}$}; \\ \&
		\node[] (n0l) {$\mathbf{3}$}; \&
		\node[] (n1l) {$\mathbf{3}$}; \&
		\node[] (n2l) {$\mathbf{4}$}; \&
		\node[] (n3l) {$\mathbf{4}$}; \&
		\node[] (n4l) {$\mathbf{6}$}; \&
		\node[] (n5l) {$\mathbf{5}$}; \&
		\node[] (n6l) {$\mathbf{5}$};\&
		\node[] (n7l) {$\mathbf{6}$};\&
		\node[] (n8l) {$\mathbf{6}$};\&
		\node[] (n9l) {$\mathbf{6}$};\&\&\&\&\&\&\&\&\&\&\&\\ \& \&
		\node[] (n1lp) {$\mathbf{3}$}; \&
		\node[] (n2lp) {$\mathbf{3}$}; \&
		\node[] (n3lp) {$\mathbf{4}$}; \&
		\node[] (n4lp) {$\mathbf{3}$}; \&
		\node[] (n5lp) {$\mathbf{3}$}; \&
		\node[] (n6lp) {$\mathbf{4}$};\&
		\node[] (n7lp) {$\mathbf{4}$};\&
		\node[] (n8lp) {$\mathbf{7}$};\&
		\node[] (n9lp) {$\mathbf{7}$};\&
		\node[] (n10lp) {$\mathbf{7}$};\&
		\node[] (n11lp) {$\mathbf{7}$};\&
		\node[] (n12lp) {$\mathbf{7}$};\&
		\node[] (n13lp) {$\mathbf{5}$};\&
		\node[] (n14lp) {$\mathbf{5}$};\&
		\node[] (n15lp) {$\mathbf{4}$};\&
		\node[] (n16lp) {$\mathbf{3}$};\&
		\node[] (n17lp) {$\mathbf{3}$};\&
		\node[] (n18lp) {$\mathbf{3}$};\&
		\node[] (n19lp) {$\mathbf{3}$};\&
		\node[] (n20lp) {$\mathbf{3}$};\\
	};
	
	\draw[thick] (g0) -- (g1) -- (g2) -- (g3) -- (g4) -- (g5) -- (g6) -- (g7) -- (g8) -- (g9) -- (g10) -- (g11) -- (g12) -- (g13) -- (g14) -- (g15) -- (g16) -- (g17) -- (g18) -- (g19) -- (g20);
	\draw[thick] (n0) -- (n1) -- (n2) -- (n3) -- (n4) -- (n5) -- (n6) -- (n7) -- (n8) -- (n9) -- (n10) -- (n11) -- (n12) -- (n13) -- (n14) -- (n15) -- (n16) -- (n17) -- (n18) -- (n19) -- (n20);
	
	\draw[thick] (g0p) -- (g1p) -- (g2p) -- (g3p) -- (g4p) -- (g5p) -- (g6p) -- (g7p) -- (g8p) -- (g9p) -- (g10p) -- (g11p) -- (g12p) -- (g13p) -- (g14p) -- (g15p) -- (g16p) -- (g17p) -- (g18p) -- (g19p) -- (g20p);
	\draw[thick] (n0p) -- (n1p) -- (n2p) -- (n3p) -- (n4p) -- (n5p) -- (n6p) -- (n7p) -- (n8p) -- (n9p) -- (n10p) -- (n11p) -- (n12p) -- (n13p) -- (n14p) -- (n15p) -- (n16p) -- (n17p) -- (n18p) -- (n19p) -- (n20p);
	
	\draw[thick] (g0l) -- (g1l) -- (g2l) -- (g3l) -- (g4l) -- (g5l) -- (g6l) -- (g7l) -- (g8l) -- (g9l) -- (g10l) -- (g11l) -- (g12l) -- (g13l) -- (g14l) -- (g15l) -- (g16l) -- (g17l) -- (g18l) -- (g19l) -- (g20l);
	\draw[thick] (n0l) -- (n1l) -- (n2lp) -- (n3l) -- (n4lp) -- (n5lp) -- (n6lp) -- (n7lp) -- (n8l) -- (n9l) -- (n10lp) -- (n11lp) -- (n12lp) -- (n13lp) -- (n14lp) -- (n15lp) -- (n16lp) -- (n17lp) -- (n18lp) -- (n19lp) -- (n20lp);
	\end{tikzpicture}}
	\caption{Application of the proposed strategies and corresponding node counts (for the circuits $G$ and $G^{\prime}$ from Ex.~\ref{ex:dd_based_ec})}
	\label{fig:strategies}
	\vspace*{-3mm}
\end{figure*}

\subsection{The Power of Simulation}\label{sec:gen_dac}

The second characteristic we are exploiting rests on the observation that simulation is much more powerful for equivalence checking of quantum circuits than for equivalence checking of classical circuits.
More precisely, in the classical realm, it is certainly possible to simulate two circuits with random inputs to obtain counterexamples in case they are not equivalent. 
However, this often does not yield the desired result. In fact, due to masking effects and the inevitable information loss introduced by many classical gates, the chance of detecting differences in the circuits within a few arbitrary simulations is greatly reduced (e.g., $x \wedge 0$ masks any difference that potentially occurs during the calculation of~$x$).
Consequently, sophisticated schemes for constraint-based stimuli generation~\cite{yuanConstraintbasedVerification2006,bergeronWritingTestbenchesUsing2006,kitchenStimulusGenerationConstrained2007,willeSMTbasedStimuliGeneration2009}, fuzzing~\cite{laeuferRFUZZCoveragedirectedFuzz2018,leDetectionHardwareTrojans2019}, etc. are employed in order to verify classical circuits.

This is significantly different in quantum computing. Here, the inherent reversibility of quantum operations dramatically reduces these effects and frequently yields situations where even small differences remain unmasked and affect entire system matrices---showing the power %
of random simulations for checking the equivalence of quantum circuits. %
Because of that, it is in general not necessary to compare the \emph{entire} system matrices---in particular when two circuits are \emph{not} equivalent and, hence, their system matrices differ from each other.
Then, rather than constructing the overall matrices~$U$ and $U^\prime$ for both computations, it is sufficient to just compare a couple of individual columns.
If any of those columns differ, the two circuits have been shown to be \emph{non-equivalent}. This is illustrated by the following example.

\begin{example}\label{ex:compilationerror}
	Consider again the circuit~$G$ and the erroneous circuit $\tilde{G}^\prime$ from Ex.~\ref{ex:noneq}. As discussed earlier, the respective system matrices $U$ and $\tilde{U}^\prime$ are shown in Fig.~\ref{fig:u} and Fig.~\ref{fig:Uprime}, respectively.
	Upon comparison of $U$ and $\tilde{U}^\prime$, the \mbox{non-equivalence} of both circuits is clearly seen.
	Moreover, since $U$ and $\tilde{U}^\prime$ differ in \emph{all} their columns, this non-equivalence can also be detected by constructing \emph{any} two columns $\ket{u_i}$ and $\ket{\tilde{u}^\prime_i}$, rather than constructing the entire %
	matrices~$U$ and~$\tilde{U}^\prime$.
	Specifically, it holds that $\mathcal{F}(\ket{u_i},\ket{\tilde{u_i}^\prime}) \approx 0.92$ for all $i$ from $0$ to $7$. %
\end{example}

While the construction of the matrix~$U$ (and accordingly of~$U^\prime$) requires expensive \mbox{matrix-matrix} multiplications $U_{m-1}\cdots U_{0}$, the construction of a single column~\ket{u_i} with \mbox{$i\in\{0,\dots,2^n-1\}$} equates to simulating $G$ with the computational basis state \ket{i} as input, i.e., performing the \mbox{matrix-vector} multiplications
\[\resizebox{0.98\linewidth}{!}{$
\vert u^{(0)}_i \rangle = U_0 \ket{i},\quad \vert u^{(j)}_i \rangle = U_{j} \cdot \vert u^{(j-1)}_i\rangle \mbox{ for } j\in\{1,\dots,m-1\}.
$}\]
If the results of those simulations (respectively yielding the $i^{\mathit{th}}$~columns $\ket{u_i} = \vert u^{(m-1)}_i\rangle$ and $\ket{u^\prime_i}= \vert u^{\prime(m^\prime-1)}_i\rangle$) differ, the two circuits have been shown to be \emph{non-equivalent}.

This constitutes an exponentially easier task than constructing the entire system matrices $U$ and $U^\prime$---although the complexity of simulation still remains exponential with respect to the number of qubits.
Regarding the complexity, creating the entire system matrices corresponds to simulating the respective circuit with all $2^n$ different computational basis states.
All this, of course, does not guarantee that any difference is indeed detected by just simulating a limited number of arbitrary computational basis states~$\ket{i}$.
But motivated by these observations, a more detailed consideration of this direction was triggered whose results are summarized in Section~\ref{sec:feasib}.
In combination with the ideas from Section~\ref{sec:gen_aspdac}, this eventually leads to the proposal of an advanced equivalence checking flow (described in detail in Section~\ref{sec:ecflow})
which, first, quickly checks for a possible non-equivalence with some simulation runs. %
If no counterexample has been obtained by this, the (highly probable) equivalence is proven afterwards.
As evaluations summarized in Section~\ref{sec:experiments} confirm, this dramatically outperforms the state of the art.

\section{Strategies for conducting $G\shortrightarrow \mathbb{I} \shortleftarrow G^\prime$}\label{sec:strategies}

Following the general ideas outlined in Section~\ref{sec:gen_aspdac} potentially allows one to conduct %
equivalence checking of quantum circuits with decision diagrams in a significantly more efficient fashion than before. 
However, to fully exploit the idea's potential, a  ``good'' strategy for how to eventually conduct $G\rightarrow\mathbb{I}\leftarrow G^{\prime}$ (i.e., when to apply gates from~$G$ and when to apply inverted gates from~$G^{\prime}$) is essential. 
In this section, we propose several promising strategies and illustrate their application. 

\subsection{Naive Strategy}\label{sec:naive}
The first strategy is motivated by the (rather \emph{naive}) assumption that a given circuit $G$ is checked against itself, i.e. $G\rightarrow\mathbb{I}\leftarrow G$. Then, obviously the best possible strategy is to alternate between applications of gates from $G$ and their respective inverse---yielding the identity function after each pair of operations. 
In case that $G\neq G^\prime$ (and, without loss of generality, assuming that $m<m^\prime$, i.e., that $G^\prime$ has more gates), this strategy 
alternates between~$G$ and~$G^{\prime}$ until all gates of~$G$ have been applied. Afterwards, the ``left-over''~gates from~$G^{\prime}$ are applied. This strategy supposedly works well if~$G$ and~$G^{\prime}$ are very similar, but obviously looses its benefits if both circuits  significantly differ in their structure (in particular if one circuit has significantly more gates than the other, i.e., if~$m\ll m^\prime$).

\begin{example}
	Consider again the two circuits $G$ and $G^{\prime}$ as discussed before in Ex.~\ref{ex:dd_based_ec}. 
	Applying the naive strategy leads to an order of gate applications as shown at the top of Fig.~\ref{fig:strategies} with the corresponding node count listed below the respective identifier.
	During this process, 
	the size of the (intermediate) decision diagrams never exceeds~$7$ nodes, while the average node count is~$4.43$.
	This is significantly less than required by the state-of-the-art approach which has a maximal node count of $9$ and an average one of $5.95$.
\end{example}

\subsection{Proportional Strategy}\label{sec:proportional}
Applying the naive strategy to circuits which, structurally, are significantly different obviously leads to an imbalance since a huge portion of ``left-over'' gates are applied---possibly neglecting the effect of staying close to the identity function. In order to avoid that, the \emph{proportional} strategy aims for a more balanced approach. To this end, first, the ratio with respect to the number of gates for both circuits is determined. Afterwards, the gates from $G$ and $G^{\prime}$ are proportionally applied according to this ratio.

\begin{example}
	Consider again the two circuits $G$ and $G^{\prime}$ as discussed before in Ex.~\ref{ex:dd_based_ec}. 
	The ratio between their gate counts is $4:16 = 1:4$. 
	Hence, applying the proportional strategy leads to an order of gate applications as shown in the second row of Fig.~\ref{fig:strategies} with the corresponding node count listed below the respective identifier.
	During this process, the size of the (intermediate) decision diagrams never exceeds~$7$ nodes, while the average node count is~$4.33$.
	For the problem at hand, this strategy constitutes near optimal performance.
\end{example}

\subsection{Look-ahead Strategy}\label{sec:lookahead}
Despite strategies considering structural elements of the given circuits for deciding the actions to be performed, also schemes based on the actual size of the (intermediate) decision diagrams may provide a good indication of how to proceed.  
Recall that the general aim is to stay as close as possible to the identity function (leading to the smallest possible decision diagram). 
Hence, the decision to 
apply a gate either from $G$ or $G^{\prime}$ can be based on which case actually leads to a smaller decision diagram. 
This is conducted by the \emph{look-ahead} scheme. 
While this potentially doubles the number of multiplications to be performed (since both alternatives have to be checked out), it may lead to smaller decision diagrams and, by this, a more efficient equivalence checking routine. 

\begin{example}
	Consider again the two circuits $G$ and $G^{\prime}$ as discussed before in Ex.~\ref{ex:dd_based_ec}.
	Applying the look-ahead strategy leads to an order of gate applications as shown in the third row of Fig.~\ref{fig:strategies}.
	As long as there are still gates left to be multiplied in \emph{both} circuits, the bottom of Fig.~\ref{fig:strategies} indicates the node count of both alternatives ($G$ on top, $G^\prime$ on the bottom). The bold line indicates which path was chosen.
	The resulting sequence of operations is exactly the one shown in Fig.~\ref{fig:g_ginverse_id} which was employed in Ex.~\ref{ex:proposed_approach}---resulting in a maximum of~$7$ nodes, while the average node count is~$4.24$.
\end{example}

Even for the small example showcased throughout this section, all proposed strategies perform significantly better in terms of maximum as well as average decision diagram size when compared to the state-of-the-art approach. 

\section{Investigating the Power of Simulation}\label{sec:feasib}

In Section~\ref{sec:gen_dac}, we illustrated the potential power of simulation for equivalence checking and argued that, in case 
two quantum circuits \mbox{$G$} and \mbox{$G^\prime$} operating on $n$ qubits are \emph{not} equivalent, even a few simulation runs will likely yield a counterexample. 
This idea triggered a more detailed consideration in which we elaborated how significantly the matrices $U$ and $U^\prime$ differ from each other in case of \mbox{non-equivalence} %
and whether this would make an incomplete coverage of the functionality feasible.
The results obtained by this consideration are summarized in this section.

To this end, we first introduce the notion of the \emph{difference} of two unitary matrices. 
Given two unitary matrices $U$ and $U^\prime$, their \emph{difference} $D$ is defined as the unitary matrix $D=U^{\dag}U^\prime$ and it holds that $U \cdot D = U^\prime$.\footnote{Similarly to picking $G^{\prime -1} \cdot G$ in Section~\ref{sec:gen_aspdac}, the particular arrangement $D=U^{\dag}U^\prime$ is just one of the four possibilities to introduce the notion of the \emph{difference} of two unitaries.}
In case both matrices are identical (i.e., the circuits are equivalent), it directly follows that \mbox{$D=\mathbb{I}$}.
One characteristic of the identity function~$\mathbb{I}$ resulting in this case is that all diagonal entries are equal to one, i.e.,~$\bra{i} U^{\dag}U^\prime \ket{i} = 1$ for $i\in\{0,\dots,2^n-1\}$, where $\ket{i}$ denotes the $i^{th}$ computational basis state.
More generally---in case of a potential relative/global phase difference between $G$ and $G^\prime$---all diagonal elements have modulus one, i.e.,~\mbox{$\vert\bra{i} U^{\dag}U^\prime \ket{i}\vert^2 = 1$}.
This expression can further be rewritten to %
\begin{align*}
1 &= \vert\bra{i} U^{\dag}U^\prime \ket{i}\vert^2 = \vert(U\ket{i})^\dag (U^\prime \ket{i})\vert^2
 = \vert\ket{u_i}^\dag \ket{u_i^\prime}\vert^2 \\&= \vert\braket{u_i}{u_i^\prime}\vert^2,
\end{align*}
where $\ket{u_i}$ and $\ket{u_i^\prime}$ denote the $i^{th}$ column of $U$ and $U^\prime$, respectively.
This essentially resembles the simulation of both circuits with the initial state $\ket{i}$ and, afterwards, calculating the fidelity $\mathcal{F}$ between the resulting states $\ket{u_i}$ and $\ket{u_i^\prime}$ (see Section~\ref{sec:considered}).  
Hence, if only one simulation yields~\mbox{$\mathcal{F}_i\coloneqq\mathcal{F}(\ket{u_i}, \ket{u_i^\prime}) \not\approx 1$}, then $\ket{i}$ proves the \mbox{non-equivalence} of  $G$ and $G^\prime$.

Now, the question is how many computational basis states~$\ket{i}$ yield $\mathcal{F}_i \not\approx 1$ for a given difference matrix $D$, i.e.,~how likely it is for an arbitrary simulation to detect possible differences.
Since the difference $D$ of both matrices is unitary itself, it may as well be interpreted as a quantum circuit $G_D$. 
For the purpose of this theoretical consideration, we assume that each gate of $G_D$ either represents a single-qubit or a (multi-)controlled operation\footnote{This does not limit the applicability of the following findings, since arbitrary single-qubit operations combined with {\sc CNOT} form a universal gate-set~\cite{nielsenQuantumComputationQuantum2010}.}.

\begin{example}\label{ex:bestcase}
	Assume that $G_D$ only consists of one (non-trivial) single-qubit operation defined by the matrix $U_s$ applied to the first of $n$ qubits. Then, the system matrix $D$ is given by
	\[
	D =  \mathbb{I}_{2^{n-1}} \otimes U_s =  
	\begin{bNiceArray}{LCR}[small] U_s & & \\ & \Ddots & \\ & & U_s 
	\end{bNiceArray}.
	\]
	The process of going from $U$ to $U^\prime$, i.e., calculating $U\cdot D$, impacts \emph{all} columns of $U$.
	Thus, an error is detected by a \emph{single} simulation with \emph{any} computational basis state.
\end{example}

Among all quantum operations, single-qubit operations posses a system matrix least similar to the identity matrix due to the tensor product structure of their corresponding system matrix. 

\begin{example}\label{ex:worstcase}
	In contrast to Ex.~\ref{ex:bestcase}, assume that $G_D$ only consists of one operation $U_s$ targeted at the first qubit and controlled by the remaining $n-1$ qubits. Then, the corresponding system matrix is given by
	\[
	D = \scalebox{0.85}{$\begin{bNiceArray}{LCCC}[small, nullify-dots]
	\mathbb{I}_{2} & &  &  \\
	& \Ddots & &\\
	& & \mathbb{I}_2 & \\
	& & & U_s
	\end{bNiceArray}$}.
	\]
	In this case, applying $D$ to $U$ only affects the last two columns of $U$. As a consequence, a maximum of two columns (out of~$2^n$) may serve as counterexamples---the worst case scenario.
\end{example}

These basic examples cover the extreme cases when it comes to the difference of two unitary matrices. In case $G_D$ exhibits no such simple structure, the analysis is more involved, e.g., generally quantum operations with \mbox{$c\in\{0,\dots,n-1\}$} controls will exhibit a difference in $2^{n-c}$~columns. Furthermore, given two operations showing a certain number of differences, the matrix product of these operations in most cases (except when cancellations occur) differs in as many columns as the maximum of both operands.

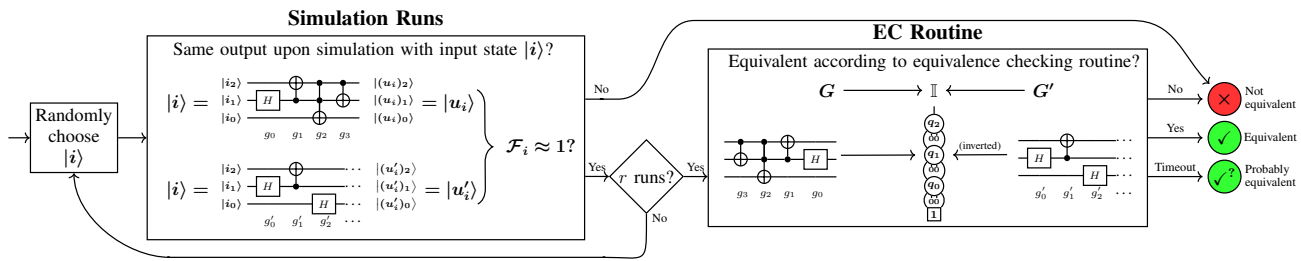
\begin{figure*}
	\centering
	\resizebox{0.95\linewidth}{!}{
	\begin{tikzpicture}
	\node[thick,draw,font=\large,rectangle, align=center] (rand) {Randomly\\choose\\$\bm{\ket{i}}$};
	
	\node[thick,draw,font=\scriptsize,rectangle, right=0.65 of rand] (sim) {
		\begin{tikzpicture}
			\node[] (G) {
				\begin{quantikz}[column sep=6pt, row sep={0.4cm,between origins}, ampersand replacement=\&]
				\lstick{$\bm{\ket{i_2}}$} \& \qw \& \targ{} \& \ctrl{2} \& \ctrl{1} \& \qw \& \rstick{$\bm{\ket{(u_i)_2}}$} \\
				\lstick{$\bm{\ket{i_1}}$}  \& \gate{H} \& \ctrl{-1} \& \ctrl{1} \&\targ{}\& \qw \& \rstick{$\bm{\ket{(u_i)_1}}$} \\
				\lstick{$\bm{\ket{i_0}}$} \& \qw  \& \qw \& \targ{} \& \qw \& \qw \& \rstick{$\bm{\ket{(u_i)_0}}$} \\
				\& \lstick[label style={xshift=0.3cm}]{$g_0$} \& \lstick[label style={xshift=0.3cm}]{$g_1$} \& \lstick[label style={xshift=0.3cm}]{$g_2$} \& \lstick[label style={xshift=0.3cm}]{$g_3$} 
				\end{quantikz}
			};
	
			\node[below=2. of G.west, anchor=west] (Gp) {
				\begin{quantikz}[column sep=6pt, row sep={0.4cm,between origins}, ampersand replacement=\&]
				\lstick{$\bm{\ket{i_2}}$} \& \qw \& \targ{} \& \qw  \& \push{\bm{\cdots}} \& \rstick{$\bm{\ket{(u_i^\prime)_2}}$} \\
				\lstick{$\bm{\ket{i_1}}$}  \& \gate{H} \& \ctrl{-1}\& \qw  \&\push{\bm{\cdots}} \& \rstick{$\bm{\ket{(u_i^\prime)_1}}$} \\
				\lstick{$\bm{\ket{i_0}}$} \& \qw  \& \qw \& \gate{H}  \& \push{\bm{\cdots}} \& \rstick{$\bm{\ket{(u_i^\prime)_0}}$}  \\
				\& \lstick[label style={xshift=0.3cm}]{$g_0^\prime$} \& \lstick[label style={xshift=0.3cm}]{$g_1^\prime$} \& \lstick[label style={xshift=0.3cm}]{$g_2^\prime$} \& \lstick[label style={xshift=0.35cm}]{$\bm{\cdots}$}
				\end{quantikz}
			};

		\node[inner sep=0pt,above=\abovecaptionskip of G.45,text width=0.51\linewidth, align=center, font=\large] {\hspace*{0.4cm}Same output upon simulation with input state $\bm{\ket{i}}$?};
		
		\node[font=\large,thick] at ($(G.west) + (-1.1,0.1)$) {$\bm{\ket{i}=}$};
		\node[font=\large,thick] at ($(Gp.west) + (-1.1,0.1)$) {$\bm{\ket{i}=}$};
		
		\node[font=\large,thick] (ui) at ($(G.east) + (-0.1,0.1)$) {$\bm{=\ket{u_i}}$};
		\node[font=\large,thick] (uip) at ($(Gp.east) + (-0.1,0.1)$) {$\bm{=\ket{u_i^\prime}}$};
		
		\draw [decorate,decoration={brace,amplitude=10pt,raise=4pt}, thick]
		($(ui.45)+(0.2,0)$) -- ($(uip.-45)+(0.15,0)$) node [midway,xshift=0.7cm, font=\large] {$\bm{\mathcal{F}_i \approx 1?}$};
		\end{tikzpicture}
	};
	\node[inner sep=0pt,above=\abovecaptionskip of sim,text width=0.4\linewidth, align=center, font=\Large] {\textbf{Simulation Runs}};

	\node[thick,draw,font=\scriptsize,rectangle, right=2.8 of sim] (ec) {
		\begin{tikzpicture}
			\node[] (G) {
				\begin{quantikz}[column sep=6pt, row sep={0.4cm,between origins}, ampersand replacement=\&]
				\& \ctrl{1} \& \ctrl{2} \& \targ{} \& \qw\& \qw \\
				\& \targ{} \& \ctrl{-1} \& \ctrl{-1} \&\gate{H}\& \qw \\
				\& \qw  \& \targ{} \& \qw \& \qw \& \qw \\
				\& \lstick[label style={xshift=0.3cm}]{$g_3$} \& \lstick[label style={xshift=0.3cm}]{$g_2$} \& \lstick[label style={xshift=0.3cm}]{$g_1$} \& \lstick[label style={xshift=0.3cm}]{$g_0$} 
				\end{quantikz}
			};
			
			\node[right=3.8 of G.east, anchor=west] (Gp) {
				\begin{quantikz}[column sep=6pt, row sep={0.4cm,between origins}, ampersand replacement=\&]
				\& \qw \& \targ{} \& \qw  \& \push{\bm{\cdots}}  \\
				\& \gate{H} \& \ctrl{-1}\& \qw  \&\push{\bm{\cdots}} \\
				\& \qw  \& \qw \& \gate{H}  \& \push{\bm{\cdots}} \\
				\& \lstick[label style={xshift=0.3cm}]{$g_0^\prime$} \& \lstick[label style={xshift=0.3cm}]{$g_1^\prime$} \& \lstick[label style={xshift=0.3cm}]{$g_2^\prime$} \& \lstick[label style={xshift=0.35cm}]{$\bm{\cdots}$}
				\end{quantikz}
			};
		
			\node[thick,draw, circle, inner sep=1pt] (q1) at ($(G.10)!0.5!(Gp.170) + (0,0)$) {$\bm{q_1}$};
			\node[thick,draw, circle, inner sep=1pt, above=0.25 of q1] (q2) {$\bm{q_2}$};
			\node[thick,draw, circle, inner sep=1pt, below=0.25 of q1] (q0) {$\bm{q_0}$};
			\node[thick,draw, rectangle, inner sep=2pt, below=0.25 of q0] (t) {$\bm{1}$};
			\draw[thick] (q2) -- ++ (0, 0.4);
			
			\draw[thick] (q2.-135) to [out=-135,in=135] (q1.135);
			\draw[thick] (q2.-105) -- ++ (0, -0.05) node[below, inner sep = 0pt] {\tiny$\bm{0}$};
			\draw[thick] (q2.-75) -- ++ (0, -0.05) node[below, inner sep = 0pt] {\tiny$\bm{0}$};		
			\draw [thick](q2.-45) to [out=-45,in=45] (q1.45);
			
			\draw[thick] (q1.-135) to [out=-135,in=135] (q0.135);
			\draw[thick] (q1.-105) -- ++ (0, -0.05) node[below, inner sep = 0pt] {\tiny$\bm{0}$};
			\draw[thick] (q1.-75) -- ++ (0, -0.05) node[below, inner sep = 0pt] {\tiny$\bm{0}$};		
			\draw[thick] (q1.-45) to [out=-45,in=45] (q0.45);
			
			\draw[thick] (q0.-135) to [out=-135,in=135] (t.135);
			\draw[thick] (q0.-105) -- ++ (0, -0.05) node[below, inner sep = 0pt] {\tiny$\bm{0}$};
			\draw[thick] (q0.-75) -- ++ (0, -0.05) node[below, inner sep = 0pt] {\tiny$\bm{0}$};
			\draw[thick] (q0.-45) to [out=-45,in=45] (t.45);
		
			\draw[->,thick, shorten >=0.2cm] (G.10) -- (q1);
			\draw[->,thick, shorten >=0.2cm] (Gp.170) -- (q1) node[midway,above,xshift=0.1cm] {(inverted)};
			
			\node[above=0.3 of q2,font=\large] (i) {$\bm{\mathbb{I}}$};
			\node[left=2 of i,font=\large] (g) {$\bm{G}$};
			\node[right=2 of i,font=\large] (gp) {$\bm{G^\prime}$};
			\draw[->,thick,shorten >=0.1cm,shorten <=0.1cm] (g) -- (i);
			\draw[->,thick,shorten >=0.1cm,shorten <=0.1cm] (gp) -- (i);
			\node[inner sep=0pt,above=\abovecaptionskip of i,text width=0.525\linewidth, align=center, font=\large] {Equivalent according to equivalence checking routine?};
		\end{tikzpicture}
	};
	\node[inner sep=0pt,above=\abovecaptionskip of ec,text width=0.4\linewidth, align=center, font=\Large] {\textbf{EC Routine}};
	
	\draw[->,thick] (ec.10) -- ++ (1.25,0) node[above,midway] {No};
	\draw[->,thick] (ec.0) -- ++ (1.25,0) node[above,midway] {Yes};
	\draw[->,thick] (ec.-10) -- ++ (1.25,0) node[above,midway] {Timeout};
	
	\node[thick,draw,circle,fill=red!80, inner sep=1pt,minimum size=0.75cm] (x) at ($(ec.10) + (1.75,0)$) {\color{black}{\large$\bm{\times}$}};
	\node[thick,draw,circle,fill=green!80, inner sep=1pt, minimum size=0.75cm] at ($(ec.0) + (1.75,0)$) {\color{black}{\large$\bm{\checkmark}$}};
	\node[thick,draw,circle,fill=green!80, inner sep=1pt, minimum size=0.75cm] at ($(ec.-10) + (1.75,0)$) {\color{black}{\large$\bm{\checkmark^?}$}};
	
	\node[inner sep=1pt, align=left] (noneq) at ($(ec.10) + (2.8,0)$) {Not\\equivalent};
	\node[inner sep=1pt, align=left] at ($(ec.0) + (2.8,0)$) {Equivalent};
	\node[inner sep=1pt, align=left] at ($(ec.-10) + (2.8,0)$) {Probably\\equivalent};
	
	\node[draw,thick,diamond,font=\large,inner sep = 1pt] at ($(sim.-10)!0.5!(ec.-170)$) (rruns) {$r$ runs?};
	\draw[->,thick] (sim.-10) -- (rruns) node[midway,above] {Yes};
	\draw[->,thick] (rruns) -- (ec.-170) node[midway,above] {Yes};
	\draw[->,thick] (rruns) -- ++(0,-1.05) node[midway,right] {No} 
	.. controls +(-90:1) and +(0:1) .. ++(-1,-0.8)
	.. controls +(0:1) and +(180:1) .. ++(-9.9,0)
	.. controls +(180:1) and +(-90:1) .. (rand.south);
	
	\draw[->,thick] (rand) -- (sim);
	\draw[->,thick] ($(rand) + (-1.5,0)$) -- (rand);
	
	\draw[->,thick, shorten >=0.1cm] (sim.10)
	.. controls +(0:1) and +(180:1) .. ($(sim.10)+(0.75,0)$) node[midway,above] {No}
	.. controls +(0:1) and +(180:1) .. ($(sim.10)+(2.5,1.8)$)
	.. controls +(0:1) and +(180:1) .. ($(sim.10)+(12.5,1.8)$)
	.. controls +(0:1) and +(135:1) .. (x.135);
	\end{tikzpicture}}
	\vspace*{-2mm}
	\caption{Proposed equivalence checking flow}
	\label{fig:flow_diagram}
	\vspace*{-6mm}
\end{figure*}

\begin{example}\label{ex:multiple_ops}
Consider a two-qubit system and two circuits $G$ and $G^\prime$ exhibiting a difference-circuit $G_D$ consisting of two gates---the first of which is a Hadamard operation $H(q_1)$ on the second qubit, while the second is a \qop{CNOT} operation~$\qop{CNOT}(q_1, q_0)$ with target on the first and control on the second qubit.
As discussed in Ex.~\ref{ex:bestcase} and Ex.~\ref{ex:worstcase}, the application of the single-qubit matrix $U_H$ affects all columns, while the application of the controlled two-qubit operation matrix $U_{\qop{CNOT}}$ only affects two of the four columns. Their combination, however, still affects \emph{all} columns of the result.
\end{example}

The gate-set provided by (current) quantum computers typically includes only (certain) single-qubit gates and a specific two qubit gate, such as the \qop{CNOT}~gate. Thus, multi-controlled quantum operations usually only arise at the most abstract algorithmic description of a quantum circuit and are then \emph{decomposed} into elementary operations from the device's \mbox{gate-set} before the circuit is mapped to the target architecture.
As a consequence, errors occurring during the design flow will typically consist of (1)~single-qubit errors, e.g., offsets in the rotation angle, or (2)~errors related to the application of \qop{CNOT} or \qop{SWAP}~gates.
In both cases, non-equivalence can be efficiently concluded by a limited number of simulations with arbitrary computational basis states.

Of course, this cannot be guaranteed---otherwise, the problem would hardly be \mbox{QMA-complete}.
However, following the above discussion and considering that comparing single columns (i.e.,~a partial consideration of the functionality) is substantially cheaper than a full functional coverage, the simulation of arbitrary computational basis states is a promising option.
Moreover, as also confirmed by the experimental evaluation in Section~\ref{sec:power}, if a counterexample was not obtained after a few simulations, this yields a highly probable estimate of the circuit's equivalence---in contrast to the classical realm, where this generally does not allow for any conclusion.

\section{Resulting Advanced\\Equivalence Checking Flow}\label{sec:ecflow}
\vspace*{-1mm}
The general ideas proposed in Section~\ref{sec:general} and elaborated in Section~\ref{sec:strategies} as well as Section~\ref{sec:feasib} complement each other in many different ways.
Trying to keep $G\shortrightarrow\mathbb{I}\shortleftarrow G^\prime$ close to the identity (see~Section~\ref{sec:gen_aspdac}) proves very %
efficient in case two circuits are indeed equivalent---provided a ``good'' strategy can be employed.
Conducting simulations with randomly chosen computational basis states $\ket{i}$ (see~Section~\ref{sec:gen_dac}) on the other hand allows to quickly detect non-equivalence even in cases where both circuits only differ slightly.
This section first describes how these findings eventually result in an advanced equivalence checking flow---followed by a discussion of the resulting methodology.
\vspace*{-3mm}
\subsection{Resulting Flow}\vspace*{-1mm}
The motivation and considerations above eventually led us to an advanced equivalence checking flow as shown in Fig.~\ref{fig:flow_diagram}. 
Here, instead of constructing and comparing the complete matrix representations of both circuits, we propose to first perform a limited number of $r\ll 2^n$ simulation runs with randomly chosen computational basis states.
Should one of those simulations yield different outputs in both circuits (i.e., a fidelity $\mathcal{F}_i \not\approx 1$), the non-equivalence of the circuits under consideration has been shown. 
If this is not the case, the equivalence checking routine $G\shortrightarrow\mathbb{I}\shortleftarrow G^\prime$ is utilized to complete the task.
Moreover, if simulation has not revealed any differences, the likelihood of the circuits being non-equivalent is significantly reduced as shown in the discussions from Section~\ref{sec:feasib} and confirmed by the evaluations in Section~\ref{sec:experiments}. 
Overall, this eventually leads to three possible outcomes: 
\begin{itemize}
	\item \emph{Not equivalent}, if any simulation run yields $\mathcal{F}_i \not\approx 1$ or if the equivalence checking routine is employed after the simulation runs and yields this result.\footnote{As a consequence of the discussions from Section~\ref{sec:feasib} it is far more likely that this result already occurs during simulation.} As confirmed by the evaluations summarized later in Section~\ref{sec:experiments}, this can be conducted very efficiently in many cases, while \mbox{state-of-the-art} equivalence checking routines require substantial runtime or even time out frequently.
	
	\item \emph{Equivalent}, if, after $r$ simulation runs, the equivalence checking routine is employed and yields that result. If this is the case, the simulation runs conducted before (which did not lead to a conclusive result) only constitute a negligible runtime overhead. As discussed earlier, this is because constructing the whole functionality is complexity-wise equivalent to simulating \emph{all} $2^n$ computational basis states and we have chosen $r\ll 2^n$.
	
	\item \emph{Timeout}, if the simulation runs did not lead to a counterexample and the equivalence checking routine was not able to complete the task within a given resource limit. If this is the case, we at least get an indication that both circuits might be equivalent (since the conducted simulations did not provide a counterexample which, according to the discussions from Section~\ref{sec:feasib}, is rather rare). Even if this does not provide a guarantee of non-equivalence, this is a stronger result than provided by the state of the art thus far, which does not allow for any kind of conclusion in this case.
\end{itemize}
\vspace*{-3mm}
\subsection{Discussion}
\vspace*{-1mm}
The resulting equivalence checking flow with the combination of using the power of simulation
together with the dedicated $G\shortrightarrow\mathbb{I}\shortleftarrow G^\prime$ scheme %
dramatically improves the state of the art. 
In contrast to constructing and comparing the whole functionality of two circuits, simulation can typically be conducted significantly faster while already providing valuable insights, i.e., a counterexample or an indication that both circuits might be equivalent.
If the results of simulation remain inconclusive, i.e., no counterexample was found, the subsequent functional comparison can be improved by utilizing the $G\shortrightarrow\mathbb{I}\shortleftarrow G^\prime$ scheme.
Even if the latter step (the actual proof) cannot be completed due to time or resource limits, a meaningful result has already been obtained after conducting the simulations---namely that the circuits are presumably equivalent---while previously proposed approaches provide no indication on the equivalence at all.

Naturally, there are cases where even single simulation runs prove too resource-consuming to conduct.
Especially in cases where simulation fails due to the immense complexity of the involved state vector, the $G\shortrightarrow\mathbb{I}\shortleftarrow G^\prime$ scheme still provides a complementary alternative. 
If a strategy can be employed which manages to keep the intermediate decision diagrams close to the identity, the whole procedure can again be conducted with a rather low memory footprint. 
The design of corresponding strategies could benefit from specific knowledge about the origin of both circuits, e.g., if $G^\prime$ is the result of compiling $G$ to a certain target device. It is left for future work to explore such \emph{application-specific} strategies based on the $G\shortrightarrow\mathbb{I}\shortleftarrow G^\prime$ scheme.

Besides that, our discussions in Section~\ref{sec:feasib} show that choosing computational basis states uniformly at random as input for the simulation part of the proposed equivalence checking flow promises high success rates in the quantum realm.
This is in stark contrast to the (simulative) verification of classical circuits, where sophisticated techniques such as constrained-based stimuli generation~\cite{yuanConstraintbasedVerification2006,bergeronWritingTestbenchesUsing2006,kitchenStimulusGenerationConstrained2007,willeSMTbasedStimuliGeneration2009}, fuzzing~\mbox{\cite{laeuferRFUZZCoveragedirectedFuzz2018,leDetectionHardwareTrojans2019}}, etc. have to be employed for such techniques to work at all.
Our experimental results (summarized in the next section) %
empirically show that such sophisticated techniques might not be that urgently needed in the quantum realm.
First results towards the study of quantum stimuli generation schemes have been obtained in~\cite{burgholzerRandomStimuliGeneration2021}.

Furthermore, previous works on equivalence checking of quantum circuits often employed several \emph{optimizations} such as template replacement%
~\cite{yamashitaFastEquivalencecheckingQuantum2010, duncanGraphtheoreticSimplificationQuantum2019}.
Potential candidates include optimizations frequently used during compilation~\cite{prasadDataStructuresAlgorithms2006,iwamaTransformationRulesDesigning2002,maslovQuantumCircuitSimplification2008,vidalUniversalQuantumCircuit2004,duncanGraphtheoreticSimplificationQuantum2019} (as reviewed in Section~\ref{sec:intro}).
Those optimization techniques may of course also be applied in the methodology proposed here---as long as the optimization itself is valid (which could be verified beforehand using the proposed methodology). 
In general, lowering the gate count of the respective circuits (or their concatenation in case of the $G\shortrightarrow\mathbb{I}\shortleftarrow G^\prime$ scheme) can be expected to improve the performance of the proposed equivalence checking flow (since less operations have to be applied).

Finally, there is no need to conduct both complementary steps (simulation and the $G\shortrightarrow\mathbb{I}\shortleftarrow G^\prime$ scheme) in a sequential fashion as proposed in Fig.~\ref{fig:flow_diagram}. 
In fact, the $r$ simulations and the \mbox{$G\shortrightarrow\mathbb{I}\shortleftarrow G^\prime$} scheme could also be started in parallel (provided sufficient memory and processor resources).
If then, any of the simulations yields a counterexample, the remaining calculations can be aborted. 
Vice versa, if the equivalence checking routine  returns ``equivalent'', any ongoing simulation runs can be stopped.

\vspace*{-2mm}
\section{Experimental Evaluation}\label{sec:experiments}
\vspace*{-1mm} 
In order to evaluate the ideas and methods proposed above, the %
advanced methodology for equivalence checking of quantum circuits %
(as summarized in Fig.~\ref{fig:flow_diagram}) has been implemented and integrated into the JKQ toolset~\cite{willeJKQJKUTools2020}.
For simulation, %
the method proposed in~\cite{zulehnerAdvancedSimulationQuantum2019} based on the decision diagram package proposed in~\cite{zulehnerHowEfficientlyHandle2019} has been used
and we fixed $r=16$, i.e., we conduct up to sixteen random simulations before opting for the equivalence checking routine if no counterexample has been obtained.
The dedicated strategies for conducting $G\shortrightarrow\mathbb{I}\shortleftarrow G^\prime$ (proposed in Section~\ref{sec:strategies}) have been implemented on top of the above decision diagram package as well. 
In this section, we summarize the obtained results. To this end, we first review the further setup, before the results themselves are covered.
\vspace*{-3mm}
\subsection{Setup}\label{sec:setup}
\vspace*{-1mm}
In order to evaluate the performance of the proposed equivalence checking flow, we used %
benchmarks frequently considered for evaluating %
quantum circuit compilers, simulators, and equivalence checking tools---taken from~\cite{willeRevLibOnlineResource2008}.
Each benchmark consists of a ``high-level'' description~$G$ (directly obtained from~\cite{willeRevLibOnlineResource2008}) as well as a ``low-level'' description $G^\prime$ generated by compiling %
the given descriptions into elementary gates and/or mapping %
them %
for a certain quantum computer architecture using IBM~Qiskit~\cite{aleksandrowiczQiskitOpensourceFramework2019}. 

The performance is compared against the \mbox{state-of-the-art} equivalence checking technique taken from~\cite{niemannQMDDsEfficientQuantum2016} (see~Ex.~\ref{ex:eq}).
All computations have been performed on a \SI{4}{\giga\hertz} \mbox{\emph{Amazon EC2 z1d}} instance running Ubuntu~18.04 with at least \SI{32}{\giga\byte} per job using \mbox{GNU Parallel}~\cite{tangeGNUParallel20182018}. A hard timeout of \SI{1}{\hour} (i.e., \SI[group-minimum-digits = 4]{3600}{\second}) was set for each run. 

In the following, we provide a representative subset of the obtained results. Additionally, the implementation of the proposed methodology (together with an archive of all benchmarks) is publicly available at \url{https://iic.jku.at/eda/research/quantum\_verification} for further evaluation.

\begin{table*}[htbp]
	\sisetup{table-text-alignment = right,table-format=>4.2, round-mode = places,round-precision = 2, group-minimum-digits = 4}
	\centering
	\caption{Equivalent benchmarks}\label{tab:eq}
	\label{tab:eqresults}
	\vspace*{-1mm}
\footnotesize
\resizebox{0.94\linewidth}{!}{
	\begin{tabular}{@{}lS[table-format=2.0]S[table-format=5.0]S[table-format=6.0]!{\qquad}S!{\qquad}S[table-format=2.0]SSSSSS[table-format=<1.3, round-precision=3]@{}}\toprule
		\multicolumn{4}{c}{Benchmark\hspace*{2.1em}} & \multicolumn{1}{c}{State-of-the-art\hspace*{1.75em}} & \multicolumn{6}{c}{Proposed Equivalence Checking Flow} \\
		\multicolumn{4}{c}{} & \multicolumn{1}{c}{EC Routine\hspace*{1.75em}} & \multicolumn{2}{c}{Simulation Runs} & \multicolumn{3}{c}{EC Routine} & \\
		\cmidrule(r{2.6em}){1-4}\cmidrule(lr{2.4em}){5-5}\cmidrule(lr){6-7}\cmidrule(lr){8-10}
		Name & $n$ & $\vert G \vert$ & $\vert G^\prime\vert$ & $t_{\mathit{sota}}\,[\si{\second}]$\hspace*{0.7em} & {\#sims} & {$t_{\mathit{sim}}\,[\si{\second}]$} & {$t_{\mathit{naive}}\,[\si{\second}]$} & {$t_{\mathit{prop}}\,[\si{\second}]$} & {$t_{\mathit{look}}\,[\si{\second}]$} & {$t_{\mathit{proposed}}\,[\si{\second}]$} & {$\frac{t_{\mathit{proposed}}}{t_{\mathit{sota}}}$} \\\midrule
		\begin{filecontents*}{./csv/journal_eq.csv}
apla\_203;22;80;53;13548;;;;;2,68345;16;; 
cm151a\_211;28;33;53;3889;;;;;1,0143;16;; 
rd84\_313;34;113;53;2295;;;;;0,654533;16;; 
mod5adder\_306;32;110;53;2348;;;;;0,655239;16;; 
sym9\_317;27;64;53;1540;;;63,6108;1484,83;0,386812;16;63,997612;
c2\_182;35;305;53;2545;;2664,79;2500,79;3081,5;0,733148;16;2501,523148;
cm163a\_213;29;39;53;3539;;496,948;183,925;503,433;0,940988;16;184,865988;
cm150a\_210;22;53;22;31595;3452,7;;0,591986;10,2431;3,43478;16;4,026766;0,001
rd73\_312;25;76;53;1426;1108,62;598,565;196,102;761,748;0,363455;16;196,465455;0,177
c2\_181;35;116;53;2708;403,797;60,5336;4,05569;62,5802;0,774232;16;4,829922;0,0119
example2\_231;16;157;53;19411;187,626;133,817;79,6223;132,26;3,30557;16;82,92787;0,4419
cu\_219;25;40;53;5031;177,145;123,996;72,1958;121,633;1,12273;16;73,31853;0,41388
C7552\_205;21;80;53;6384;158,3;93,9561;80,3795;94,497;1,25619;16;81,63569;0,5157
add6\_196;19;229;19;262179;150,562;79,487;13,8531;78,6219;25,9238;16;39,7769;0,2642
dk17\_224;21;49;53;6650;105,189;39,7722;36,5605;39,7756;1,21547;16;37,77597;0,3591
mlp4\_245;16;131;53;14513;79,0592;33,1327;22,6862;34,9538;2,41026;16;25,09646;0,3174
alu1\_198;20;32;53;1377;13,1852;8,43817;4,02241;7,23075;0,363671;16;4,386081;0,3328
5xp1\_194;17;85;53;5487;10,7763;5,06902;4,59428;5,15538;1,10245;16;5,69673;0,5287
pcler8\_248;21;22;53;1522;3,44161;2,06451;1,65721;1,99484;0,384707;16;2,041917;0,593
dk27\_225;18;24;53;1208;2,83136;1,82489;1,47786;1,48046;0,305863;16;1,783723;0,62897
\end{filecontents*}
\csvreader[column count=13, no head, separator=semicolon]{./csv/journal_eq.csv}
		{1=\name, 2=\qubitsg, 3=\ng, 4=\qubitsgp, 5=\ngp, 6=\tref, 7=\tnaive, 8=\tprop, 9=\tlook, 10=\tsim, 11=\nsims, 12=\tflow, 13=\impr}
		{\name &
			\qubitsg &
			\ng &
			\ngp & 
			{\ifthenelse{\equal{\tref}{}}{\num{>3600.00}}{\num{\tref}}}\hspace*{0.7em} & 
			{\ifthenelse{\equal{\tref}{}\and\equal{\tnaive}{}\and\equal{\tprop}{}\and\equal{\tlook}{}}{\bfseries \num{\nsims}}{\num{\nsims}}} & 
			{\ifthenelse{\equal{\tref}{}\and\equal{\tnaive}{}\and\equal{\tprop}{}\and\equal{\tlook}{}}{\bfseries \num{\tsim}}{\num{\tsim}}} & 
			{\ifthenelse{\equal{\tnaive}{}}{\num{>3600.00}}{\ifthenelse{\equal{\tref}{}\or\lengthtest{\tnaive pt<\tref pt}}{\bfseries\num{\tnaive}}{\num{\tnaive}}}} & 
			{\ifthenelse{\equal{\tprop}{}}{\num{>3600.00}}{\ifthenelse{\equal{\tref}{}\or\lengthtest{\tprop pt<\tref pt}}{\bfseries\num{\tprop}}{\num{\tprop}}}} & 
			{\ifthenelse{\equal{\tlook}{}}{\num{>3600.00}}{\ifthenelse{\equal{\tref}{}\or\lengthtest{\tlook pt<\tref pt}}{\bfseries\num{\tlook}}{\num{\tlook}}}} &
			{\ifthenelse{\equal{\tflow}{}}{\num{>3600.00}}{\ifthenelse{\equal{\tref}{}\or\lengthtest{\tflow pt<\tref pt}}{\bfseries\num{\tflow}}{\num{\tflow}}}} &
			{\ifthenelse{\equal{\tflow}{}}{\bfseries -}{\ifthenelse{\equal{\tref}{}}{\bfseries -}{\num{\impr}}}} \cr}		
		\\[-\normalbaselineskip]\bottomrule
	\end{tabular}}\\\vspace{1mm}
{\scriptsize $n$: Number of qubits \hspace*{0.4cm} \emph{$\vert G \vert$}: Gate count of $G$ \hspace*{0.4cm} \emph{$\vert G^\prime \vert$}: Gate count of $G^\prime$  \hspace*{0.4cm} $t_{\mathit{sota}}$: Runtime of state-of-the-art EC routine \hspace*{0.4cm} \#sims: Number of simulation runs\\ $t_{\mathit{sim}}$: Runtime of simulation runs \hspace*{0.4cm}$t_{\mathit{naive}}/t_{\mathit{prop}}/t_{\mathit{look}}$: Runtime of EC routine \hspace*{0.45cm}$t_{\mathit{proposed}}$: Runtime of proposed EC flow}\vspace*{-5mm}
\end{table*}

\vspace*{-2mm}
\subsection{Showing Equivalence}\label{sec:eq}
\vspace*{-1mm}
In a first series of evaluations, we considered cases where the two given circuits $G$ and $G^\prime$ are equivalent.
Table~\ref{tab:eq} provides the obtained results.
The first columns describe the benchmarks themselves (i.e., their name, number~$n$ of qubits, as well as the gate count $\vert G\vert$ and $\vert G^\prime\vert$), followed by the runtime $t_{\mathit{sota}}$ (in CPU seconds) for the \mbox{state-of-the-art} approach. Then, the performance of the proposed flow is listed---split into the runtime $t_{\mathit{sim}}$ for the simulation runs as well as the runtime of the individual $G\shortrightarrow\mathbb{I}\shortleftarrow G^\prime$ strategies. Finally, the last columns list the total runtime $t_{\mathit{proposed}}$ of the proposed scheme and its relative improvement compared to the state-of-the-art routine\footnote{The total runtime $t_{\mathit{proposed}}$ is calculated as the sum of the simulations' runtime and the runtime of the proportional scheme which, as discussed next, turns out to be the most efficient scheme, i.e., \mbox{$t_{\mathit{proposed}} = t_{\mathit{sim}} + t_{\mathit{prop}}$}.}.

From the results it can be seen that, in the majority of cases, the dedicated strategies proposed in this work reduce the equivalence checking time down to a half or a third compared the state-of-the-art routine. Especially the \emph{proportional} strategy performs rather well in this regard---sometimes leading to magnitudes of improvement.
Furthermore, the additionally conducted simulations only lead to a negligible run-time overhead to the overall flow. Hence, although they cannot prove equivalence, they also do not ``hurt'' much. On the contrary, as discussed in Section~\ref{sec:ecflow} and further confirmed by the following evaluations, they often provide indications of the circuits' equivalence---in particular for cases %
where no decisive answer could be obtained by any equivalence checking routine due to timeouts (e.g., for the topmost four benchmarks in Table~\ref{tab:eq}), this is better than no result at all.

\begin{table*}[htbp]
	\sisetup{table-text-alignment = right,table-format=>4.2, round-mode = places,round-precision = 2, group-minimum-digits = 4}
	\centering
	\caption{Non-equivalent benchmarks}
	\label{tab:noneqresults}\vspace*{-1mm}
	\footnotesize
	\resizebox{0.95\linewidth}{!}{
	\begin{subtable}[t]{0.99\linewidth}
		\caption{Removed 1 random gate}\label{tab:non-eq1}\vspace*{-1mm}
		\centering
		\begin{tabular}{@{}lS[table-format=2.0]S[table-format=5.0]S[table-format=6.0]!{\qquad}S!{\qquad}S[table-format=2.0]SSSSS@{}}\toprule
			\multicolumn{4}{c}{Benchmark\hspace*{2.1em}} & \multicolumn{1}{c}{State-of-the-art\hspace*{1.75em}} & \multicolumn{6}{c}{Proposed Equivalence Checking Flow} \\
			\multicolumn{4}{c}{} & \multicolumn{1}{c}{EC Routine\hspace*{1.75em}} & \multicolumn{2}{c}{Simulation Runs} & \multicolumn{3}{c}{EC Routine} & \\
			\cmidrule(r{2.6em}){1-4}\cmidrule(lr{2.4em}){5-5}\cmidrule(lr){6-7}\cmidrule(lr){8-10}
			Name & $n$ & $\vert G \vert$ & $\vert G^\prime\vert$ & $t_{\mathit{sota}}\,[\si{\second}]$\hspace*{0.7em} & {\#sims} & {$t_{\mathit{sim}}\,[\si{\second}]$} & {$t_{\mathit{naive}}\,[\si{\second}]$} & {$t_{\mathit{prop}}\,[\si{\second}]$} & {$t_{\mathit{look}}\,[\si{\second}]$} & {$t_{\mathit{proposed}}\,[\si{\second}]$} \\\midrule
			\begin{filecontents*}{./csv/journal_remove_1.csv}
apla\_203;22;80;53;13547;;;;;0,154034;1;0,154034
cm151a\_211;28;33;53;3888;;;;;0,120973;2;0,120973 
rd84\_313;34;113;53;2294;;;;;0,156159;4;0,156159 
mod5adder\_306;32;110;53;2347;;;;;0,167947;4;0,167947
c2\_182;35;305;53;2544;;;;;0,0476003;1;0,0476003
cm150a\_210;22;53;22;31594;;;;;0,621557;1;0,621557
rd73\_312;25;76;53;1425;;;387,474;;0,0218102;1;0,0218102
sym9\_317;27;64;53;1539;;2221,73;69,2159;1763,16;0,0232364;1;0,0232364
cm163a\_213;29;39;53;3538;;777,167;348,291;801,111;0,173455;3;0,173455
cu\_219;25;40;53;5030;;203,893;136,736;205,459;0,0853663;1;0,0853663
c2\_181;35;116;53;2707;899,189;121,084;32,6747;115,355;0,0453054;1;0,0453054
example2\_231;16;157;53;19410;194,51;176,581;112,937;167,36;0,183776;1;0,183776
C7552\_205;21;80;53;6383;160,897;99,0544;80,8343;93,6787;0,230203;3;0,230203
add6\_196;19;229;19;262178;143,052;72,2507;12,3072;70,555;0,328875;1;0,328875
dk17\_224;21;49;53;6649;106,033;41,8881;36,7462;42,6166;0,840489;11;0,840489
mlp4\_245;16;131;53;14512;80,9652;33,806;23,3513;33,0157;2,35855;16;25,70985
alu1\_198;20;32;53;1376;42,556;21,8185;11,2691;23,7264;0,0257131;1;0,0257131
5xp1\_194;17;85;53;5486;11,4334;5,61134;4,9569;5,61006;0,0656292;1;0,0656292
pcler8\_248;21;22;53;1521;4,41033;4,02751;2,96356;3,67746;0,189951;8;0,189951
dk27\_225;18;24;53;1207;3,44772;1,7444;1,65074;1,7851;0,0186772;1;0,0186772
\end{filecontents*}
\csvreader[column count=12, no head, separator=semicolon]{./csv/journal_remove_1.csv}
			{1=\name, 2=\qubitsg, 3=\ng, 4=\qubitsgp, 5=\ngp, 6=\tref, 7=\tnaive, 8=\tprop, 9=\tlook, 10=\tsim, 11=\nsims, 12=\tflow}
			{\name &
				\qubitsg &
				\ng &
				\ngp & 
				{\ifthenelse{\equal{\tref}{}}{\num{>3600.00}}{\num{\tref}}}\hspace*{0.7em} & 
				{\ifthenelse{\lengthtest{\nsims pt < 16 pt}}{\bfseries \num{\nsims}}{\num{\nsims}}} & 
				{\ifthenelse{\lengthtest{\nsims pt < 16 pt}}{\bfseries \num{\tsim}}{\num{\tsim}}} & 
				{\ifthenelse{\lengthtest{\nsims pt < 16 pt}}{\bfseries -}{\ifthenelse{\equal{\tnaive}{}}{\num{>3600.00}}{\ifthenelse{\equal{\tref}{}\or\lengthtest{\tnaive pt<\tref pt}}{\bfseries\num{\tnaive}}{\num{\tnaive}}} }} & 
				{\ifthenelse{\lengthtest{\nsims pt < 16 pt}}{\bfseries -}{\ifthenelse{\equal{\tprop}{}}{\num{>3600.00}}{\ifthenelse{\equal{\tref}{}\or\lengthtest{\tprop pt<\tref pt}}{\bfseries\num{\tprop}}{\num{\tprop}}}}} & 
				{\ifthenelse{\lengthtest{\nsims pt < 16 pt}}{\bfseries -}{\ifthenelse{\equal{\tlook}{}}{\num{>3600.00}}{\ifthenelse{\equal{\tref}{}\or\lengthtest{\tlook pt<\tref pt}}{\bfseries\num{\tlook}}{\num{\tlook}}}}} &
				{\ifthenelse{\equal{\tflow}{}}{\num{>3600.00}}{\ifthenelse{\equal{\tref}{}\or\lengthtest{\tflow pt<\tref pt}}{\bfseries\num{\tflow}}{\num{\tflow}}}} \cr}		
			\\[-\normalbaselineskip]\bottomrule
		\end{tabular}	%
	\end{subtable}}\\\vspace{0.5mm}
{\scriptsize $n$: Number of qubits \hspace*{0.4cm} \emph{$\vert G \vert$}: Gate count of $G$ \hspace*{0.4cm} \emph{$\vert G^\prime \vert$}: Gate count of $G^\prime$  \hspace*{0.4cm} $t_{\mathit{sota}}$: Runtime of state-of-the-art EC routine \hspace*{0.4cm} \#sims: Number of simulation runs\\ $t_{\mathit{sim}}$: Runtime of simulation runs \hspace*{0.4cm}$t_{\mathit{naive}}/t_{\mathit{prop}}/t_{\mathit{look}}$: Runtime of EC routine \hspace*{0.45cm}$t_{\mathit{proposed}}$: Runtime of proposed EC flow}\hfill\\\vspace{-0.5mm}
	\resizebox{0.95\linewidth}{!}{
	\begin{subtable}[t]{0.99\linewidth}
		\caption{Removed 3 random gates}\label{tab:non-eq3}\vspace*{-1mm}
		\centering
		\begin{tabular}{@{}lS[table-format=2.0]S[table-format=5.0]S[table-format=6.0]!{\qquad}S!{\qquad}S[table-format=2.0]SSSSS@{}}\toprule
			\multicolumn{4}{c}{Benchmark\hspace*{2.1em}} & \multicolumn{1}{c}{State-of-the-art\hspace*{1.75em}} & \multicolumn{6}{c}{Proposed Equivalence Checking Flow} \\
			\multicolumn{4}{c}{} & \multicolumn{1}{c}{EC Routine\hspace*{1.75em}} & \multicolumn{2}{c}{Simulation Runs} & \multicolumn{3}{c}{EC Routine} & \\
			\cmidrule(r{2.6em}){1-4}\cmidrule(lr{2.4em}){5-5}\cmidrule(lr){6-7}\cmidrule(lr){8-10}
			Name & $n$ & $\vert G \vert$ & $\vert G^\prime\vert$ & $t_{\mathit{sota}}\,[\si{\second}]$\hspace*{0.7em} & {\#sims} & {$t_{\mathit{sim}}\,[\si{\second}]$} & {$t_{\mathit{naive}}\,[\si{\second}]$} & {$t_{\mathit{prop}}\,[\si{\second}]$} & {$t_{\mathit{look}}\,[\si{\second}]$} & {$t_{\mathit{proposed}}\,[\si{\second}]$} \\\midrule
			\begin{filecontents*}{./csv/journal_remove_3.csv}
apla\_203;22;80;53;13545;;;;;0,500337;3;0,500337 
cm151a\_211;28;33;53;3886;;;;;0,0841264;1;0,0841264 
rd84\_313;34;113;53;2292;;;;;0,0383307;1;0,0383307 
mod5adder\_306;32;110;53;2345;;;;;0,0394787;1;0,0394787 
c2\_182;35;305;53;2542;;;;;0,0444021;1;0,0444021 
sym9\_317;27;64;53;1537;;;;;0,0265867;1;0,0265867 
cm163a\_213;29;39;53;3536;;;;;0,182445;3;0,182445 
c2\_181;35;116;53;2705;;;;;0,052069;1;0,052069 
C7552\_205;21;80;53;6381;;;;;0,120788;1;0,120788 
add6\_196;19;229;19;262176;;;;;1,65598;1;1,65598 
cu\_219;25;40;53;5028;;3437,91;2730,18;3333,57;0,156168;2;0,156168 
rd73\_312;25;76;53;1423;;1091,89;340,638;896,263;0,022;1;0,022 
cm150a\_210;22;53;22;31592;2982,74;;9,12411;46,4413;0,191179;1;0,191179 
example2\_231;16;157;53;19408;1215,63;1061,68;789,109;1062,04;0,258627;1;0,258627 
dk17\_224;21;49;53;6647;327,618;267,315;240,882;;0,0741739;1;0,0741739 
mlp4\_245;16;131;53;14510;185,999;89,5472;66,0627;88,9556;0,137896;1;0,137896 
alu1\_198;20;32;53;1374;84,761;77,2956;45,6233;70,2064;0,028267;1;0,028267 
5xp1\_194;17;85;53;5484;21,5236;13,2903;10,7306;13,0347;0,0760485;1;0,0760485 
dk27\_225;18;24;53;1205;5,22243;2,97696;2,50767;2,89753;0,0181744;1;0,0181744 
pcler8\_248;21;22;53;1519;4,02569;2,41588;1,54017;2,20446;0,0236669;1;0,0236669 
\end{filecontents*}
\csvreader[column count=12, no head, separator=semicolon]{./csv/journal_remove_3.csv}
			{1=\name, 2=\qubitsg, 3=\ng, 4=\qubitsgp, 5=\ngp, 6=\tref, 7=\tnaive, 8=\tprop, 9=\tlook, 10=\tsim, 11=\nsims, 12=\tflow}
			{\name &
				\qubitsg &
				\ng &
				\ngp & 
				{\ifthenelse{\equal{\tref}{}}{\num{>3600.00}}{\num{\tref}}}\hspace*{0.7em} & 
				{\ifthenelse{\lengthtest{\nsims pt < 16 pt}}{\bfseries \num{\nsims}}{\num{\nsims}}} & 
				{\ifthenelse{\lengthtest{\nsims pt < 16 pt}}{\bfseries \num{\tsim}}{\num{\tsim}}} & 
				{\ifthenelse{\lengthtest{\nsims pt < 16 pt}}{\bfseries -}{\ifthenelse{\equal{\tnaive}{}}{\num{>3600.00}}{\ifthenelse{\equal{\tref}{}\or\lengthtest{\tnaive pt<\tref pt}}{\bfseries\num{\tnaive}}{\num{\tnaive}}} }} & 
				{\ifthenelse{\lengthtest{\nsims pt < 16 pt}}{\bfseries -}{\ifthenelse{\equal{\tprop}{}}{\num{>3600.00}}{\ifthenelse{\equal{\tref}{}\or\lengthtest{\tprop pt<\tref pt}}{\bfseries\num{\tprop}}{\num{\tprop}}}}} & 
				{\ifthenelse{\lengthtest{\nsims pt < 16 pt}}{\bfseries -}{\ifthenelse{\equal{\tlook}{}}{\num{>3600.00}}{\ifthenelse{\equal{\tref}{}\or\lengthtest{\tlook pt<\tref pt}}{\bfseries\num{\tlook}}{\num{\tlook}}}}} &
				{\ifthenelse{\equal{\tflow}{}}{\num{>3600.00}}{\ifthenelse{\equal{\tref}{}\or\lengthtest{\tflow pt<\tref pt}}{\bfseries\num{\tflow}}{\num{\tflow}}}} \cr}		
			\\[-\normalbaselineskip]\bottomrule
		\end{tabular}	%
	\end{subtable}}\\\vspace{0.5mm}
{\scriptsize $n$: Number of qubits \hspace*{0.4cm} \emph{$\vert G \vert$}: Gate count of $G$ \hspace*{0.4cm} \emph{$\vert G^\prime \vert$}: Gate count of $G^\prime$  \hspace*{0.4cm} $t_{\mathit{sota}}$: Runtime of state-of-the-art EC routine \hspace*{0.4cm} \#sims: Number of simulation runs\\ $t_{\mathit{sim}}$: Runtime of simulation runs \hspace*{0.4cm}$t_{\mathit{naive}}/t_{\mathit{prop}}/t_{\mathit{look}}$: Runtime of EC routine \hspace*{0.45cm}$t_{\mathit{proposed}}$: Runtime of proposed EC flow}\vspace*{-5mm}
\end{table*}

\vspace*{-2mm}
\subsection{Showing Non-equivalence}\label{sec:noneq}
\vspace*{-1mm}
In a second series of evaluations, we considered cases where the two given circuits $G$ and $G^\prime$ are \emph{not} equivalent.
To this end, we injected errors into each benchmark by removing either one or a total of three random gates from the corresponding circuit~$G^\prime$.
This led to benchmarks, where the difference between $G$ and $G^\prime$ is difficult to detect (usually, differences caused by errors in the compilation flow would have much more severe effects). 
The respectively obtained results are provided in Table~\ref{tab:noneqresults}.

Using the state-of-the-art approach, detecting these differences indeed causes substantial runtimes---or cannot be completed within the given time limit at all.
In contrast, the proposed methodology can determine the non-equivalence for \emph{all} benchmarks in just a matter of seconds or even less. 
Moreover, in the vast majority of cases, simulation alone is capable of showing the non-equivalence---in more or less negligible run-time. Only in the case of \emph{mlp4\_245} (and only assuming the removal of a single gate, i.e., a most subtle difference), an actual equivalence checking routine is needed. %
Here, the $G\shortrightarrow\mathbb{I}\shortleftarrow G^\prime$ strategies can finish the job in feasible runtime. 

On top of that, the results also show that it is indeed rather unlikely that differences get masked so that they cannot be detected by simulation. While this indeed happens in case just one gate is removed (as reported in Table~\ref{tab:non-eq1} for benchmark \emph{mlp4\_245}), not a single such instance exists in our evaluation if three gates are removed (which still constitute rather subtle differences compared to the effect an actual error in the compilation flow would have on $G^\prime$). In fact, Table~\ref{tab:non-eq3}  shows that, in the majority of cases, a single simulation run (and never more than 3 runs) are necessary to detect the error---clearly demonstrating the power of simulation. 

\begin{table*}[htbp]
	\sisetup{table-text-alignment = right,table-format=>4.2, round-mode = places,round-precision = 2, group-minimum-digits = 4}
	\centering
	\caption{Simulation runtime and success probability}
	\label{tab:simulation1}\vspace*{-1mm}
	\footnotesize
	\resizebox{0.95\linewidth}{!}{
	\begin{tabular}{@{}lS[table-format=2.0]S[table-format=5.0]S[table-format=6.0]S[table-format=3.0]*{2}{!{\qquad}S[table-format=2.0, round-integer-to-decimal]SSS[table-format=1.2, round-integer-to-decimal]}@{}}\toprule
		\multicolumn{5}{c}{Benchmark\hspace*{2.6em}} & \multicolumn{4}{c}{Removed $1$ random gate\hspace*{2.6em}} & \multicolumn{4}{c}{Removed $3$ random gates}\\
		\cmidrule(r{2.6em}){1-5}\cmidrule(lr{2.6em}){6-9}\cmidrule(l){10-13}
		Name & $n$ & $\vert G \vert$ & $\vert G^\prime\vert$ & {\#inst} &
		{$\varnothing$\#sims} &
		{$\varnothing t_{\mathit{sim}}\,[\si{\second}]$} &
		{$\max t_{\mathit{sim}}\,[\si{\second}]$} &
		{\bfseries $p_{\mbox{success}}$} &
		{$\varnothing$\#sims} &
		{$\varnothing t_{\mathit{sim}}\,[\si{\second}]$} &
		{$\max t_{\mathit{sim}}\,[\si{\second}]$} &
		{\bfseries $p_{\mbox{success}}$}
		\\\midrule
		\begin{filecontents*}{./csv/journal_simulation.csv}
apla\_203;29;80;13548;100;4,4;0,624231;2,45822;0,83;1,33;0,229946;1,4265;1
cm151a\_211;37;33;3889;100;5;0,242781;0,871795;0,81;1,35;0,0677343;0,631198;1
rd84\_313;37;113;2295;100;3,56;0,12875;0,625149;0,86;1,07;0,0373478;0,148606;1
mod5adder\_306;37;110;2348;100;2,86;0,102941;0,615785;0,9;1,11;0,0416377;0,194489;1
sym9\_317;28;64;1540;100;4,1;0,0900176;0,427922;0,84;1,34;0,0315444;0,370351;0,99
c2\_182;37;305;2545;100;2,31;0,0949209;0,680905;0,97;1,14;0,0467347;0,139857;1
cm163a\_213;37;39;3539;100;4,78;0,241884;0,823843;0,84;1,36;0,0716201;0,302533;1
cm150a\_210;22;53;31595;100;4,2;0,755547;3,08683;0,82;1,17;0,211253;0,690419;1
rd73\_312;28;76;1426;100;3,55;0,0749055;0,383983;0,88;1,11;0,0242495;0,083124;1
c2\_181;37;116;2708;100;3,38;0,138959;0,691895;0,89;1,22;0,051626;0,662027;0,99
example2\_231;24;157;19411;100;4,98;0,954128;3,60974;0,81;1,66;0,312988;2,98557;0,99
cu\_219;33;40;5031;100;5,01;0,289979;0,971635;0,8;2,15;0,134894;1,05678;0,96
C7552\_205;28;80;6384;100;4,11;0,279072;1,12594;0,88;1,29;0,125009;0,5338;1
add6\_196;19;229;262179;100;4,27;6,05514;24,5287;0,81;1,58;2,532;24,406;0,98
dk17\_224;28;49;6650;100;4,75;0,306084;1,08578;0,82;1,45;0,108356;0,989306;0,99
mlp4\_245;24;131;14513;100;3,49;0,489672;2,33724;0,91;1,22;0,192815;0,668141;1
alu1\_198;28;32;1377;100;3,82;0,0808198;0,428277;0,86;1,11;0,0239349;0,178237;1
5xp1\_194;28;85;5487;100;3,98;0,237637;1,02887;0,87;1,26;0,0760031;0,256983;1
pcler8\_248;28;22;1522;100;5,29;0,137511;0,879738;0,8;1,41;0,0330226;0,365201;0,99
dk27\_225;24;24;1208;100;4,15;0,080505;0,663896;0,84;1,2;0,0278955;0,142566;1
\end{filecontents*}
\csvreader[column count=13, no head, separator=semicolon]{./csv/journal_simulation.csv}
		{1=\name, 2=\qubitsg, 3=\ng, 4=\ngp, 5=\inst, 6=\simsone, 7=\tone, 8=\maxtone, 9=\pone, 10=\simsthree, 11=\tthree, 12=\maxtthree, 13=\pthree}
		{\name &
			\qubitsg &
			\ng &
			\ngp & 
			\inst &
			\simsone &
			\tone &
			\maxtone &
			{\bfseries \num{\pone}} &
			\simsthree &
			\tthree &
			\maxtthree &
			{\bfseries \num{\pthree}}\cr}
		\\[-\normalbaselineskip]\bottomrule
	\end{tabular}}\\\vspace{0.5mm}
{\scriptsize $n$: Number of qubits \hspace*{0.45cm} \emph{$\vert G \vert$}: Gate count of $G$ \hspace*{0.45cm} \emph{$\vert G^\prime \vert$}: Original gate count of $G^\prime$  \hspace*{0.45cm} \#inst: Considered instances  \hspace*{0.45cm} $\varnothing$\#sims: Average number of simulation runs  \\  $\varnothing t_{\mathit{sim}}$: Average runtime of simulation runs \hspace*{0.45cm}  $\max t_{\mathit{sim}}$: Maximum runtime of simulation runs \hspace*{0.45cm} {\bfseries $p_{\mbox{success}}$}: Success probability}\vspace*{-5mm}
\end{table*}

\subsection{The Power of Simulation}\label{sec:power}

In order to further evaluate the power of simulation, we finally conducted a third series of evaluations in which 
the success probability of the simulation part was evaluated in detail. 
To this end, not just a single erroneous instance has been considered for each benchmark, but a hundred different instances with either one or three gates removed have been created. Afterwards, we again executed %
$r=16$ simulation runs to check %
whether this shows the respective non-equivalence. 
Table~\ref{tab:simulation1} provides the obtained results, i.e., the average number of simulations, the average and the maximum simulation runtime, as well as the success probability, i.e., the probability that non-equivalence was detected within sixteen simulations. %

In cases where just a single gate was removed from the circuit (which, thus far, was really hard, if not impossible, to detect), the success probability is \SI{85.2}{\percent} on average. %
Already this provides a rather strong argument that, when checking the equivalence of quantum circuits, simulation frequently is sufficient (in contrast to the classical realm, where masking effects frequently make simulation insufficient). Moreover, removing just two more gates (still constituting a rather subtle difference compared to the effect an actual error in the compilation flow may have on $G^\prime$) results in an average success probability which is very close to \SI{100}{\percent}. %
Additionally considering that the runtime for that  is almost always negligible, this clearly shows the power of simulation to either quickly show the non-equivalence of two circuits or, at least, provide an indication of equivalence.

\vspace*{-2mm}
\section{Conclusions}\label{sec:conclusion}
In this paper, we proposed an advanced methodology for equivalence checking of quantum circuits which takes the different paradigms of quantum computing not as burden, but as an opportunity. 
If a ``good'' strategy for keeping $G\shortrightarrow\mathbb{I}\shortleftarrow G^\prime$ close to the identity is available, obtaining the complete proof of two circuit's equivalence can be dramatically accelerated. 
Furthermore, %
we showed that, in contrast to classical computing, inherent characteristics of quantum circuits allow for the conclusion that simulations provide an indication on whether two circuits are equivalent---even in the case of rather minor differences.  
Experimental evaluations confirmed that the resulting methodology allows one to check the equivalence or \mbox{non-equivalence} 
magnitudes faster than ever before---and, in many cases, by simulation only.

Specifically, strategies for keeping $G\shortrightarrow\mathbb{I}\shortleftarrow G^\prime$ close to the identity may be derived from 
utilizing information about how a given circuit~$G$ is compiled to a resulting implementation~$G^\prime$. 
First studies on verifying the results of IBM's quantum circuit compilation flow Qiskit~\cite{aleksandrowiczQiskitOpensourceFramework2019} employing the proposed methodology have been conducted in~\cite{burgholzerVerifyingResultsIBM2020}.
In contrast to the classical realm, this could eventually make verifying the results of sophisticated design flows feasible in general.
An implementation of the proposed equivalence checking flow is publicly available at \url{https://iic.jku.at/eda/research/quantum_verification/}.

\section*{Acknowledgments}

This work has partially been supported by the LIT Secure and Correct Systems Lab funded by the State of Upper Austria as well as by the BMK, BMDW, and the State of Upper Austria in the frame of the COMET program (managed by the FFG).

\vspace*{-2mm}
\printbibliography

\begin{IEEEbiography}
	[{\includegraphics[width=1in,height=1.25in,clip,keepaspectratio]{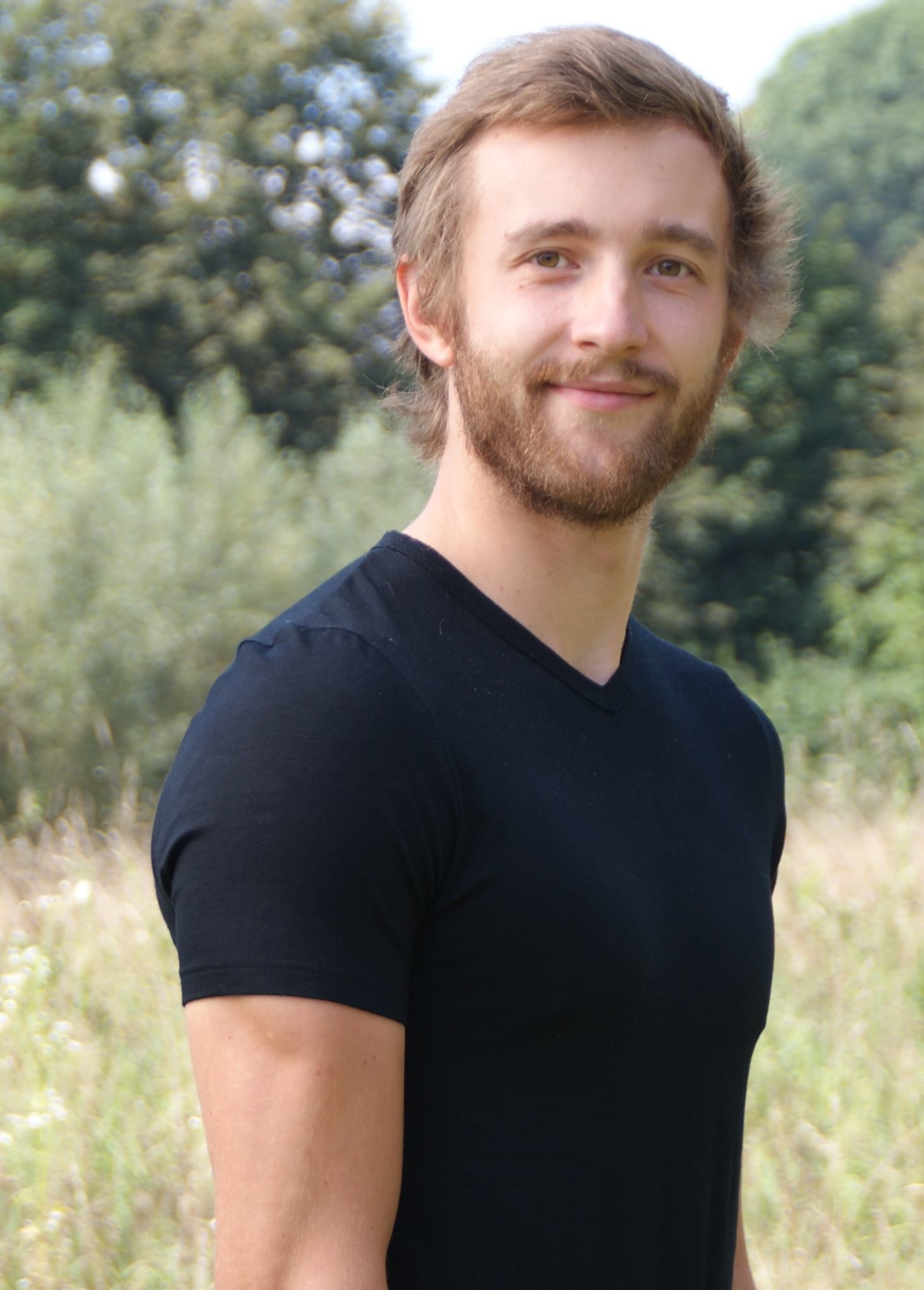}}]{Lukas Burgholzer}
	Lukas Burgholzer (S’19) received his Master's degree in industrial mathematics (2018) and Bachelor's degree in computer science (2019) from the Johannes Kepler University Linz, Austria.
	He is currently a Ph.D. student at the Institute for Integrated Circuits at the Johannes Kepler University Linz, Austria. 
	His research interests include design automation for quantum computing---currently focusing on efficient and correct compilation. In these areas, he has published several papers on international conferences such as ASP-DAC, DAC, ICCAD, and QCE.
\end{IEEEbiography}

\begin{IEEEbiography}
	[{\includegraphics[width=1in,height=1.25in,clip,keepaspectratio]{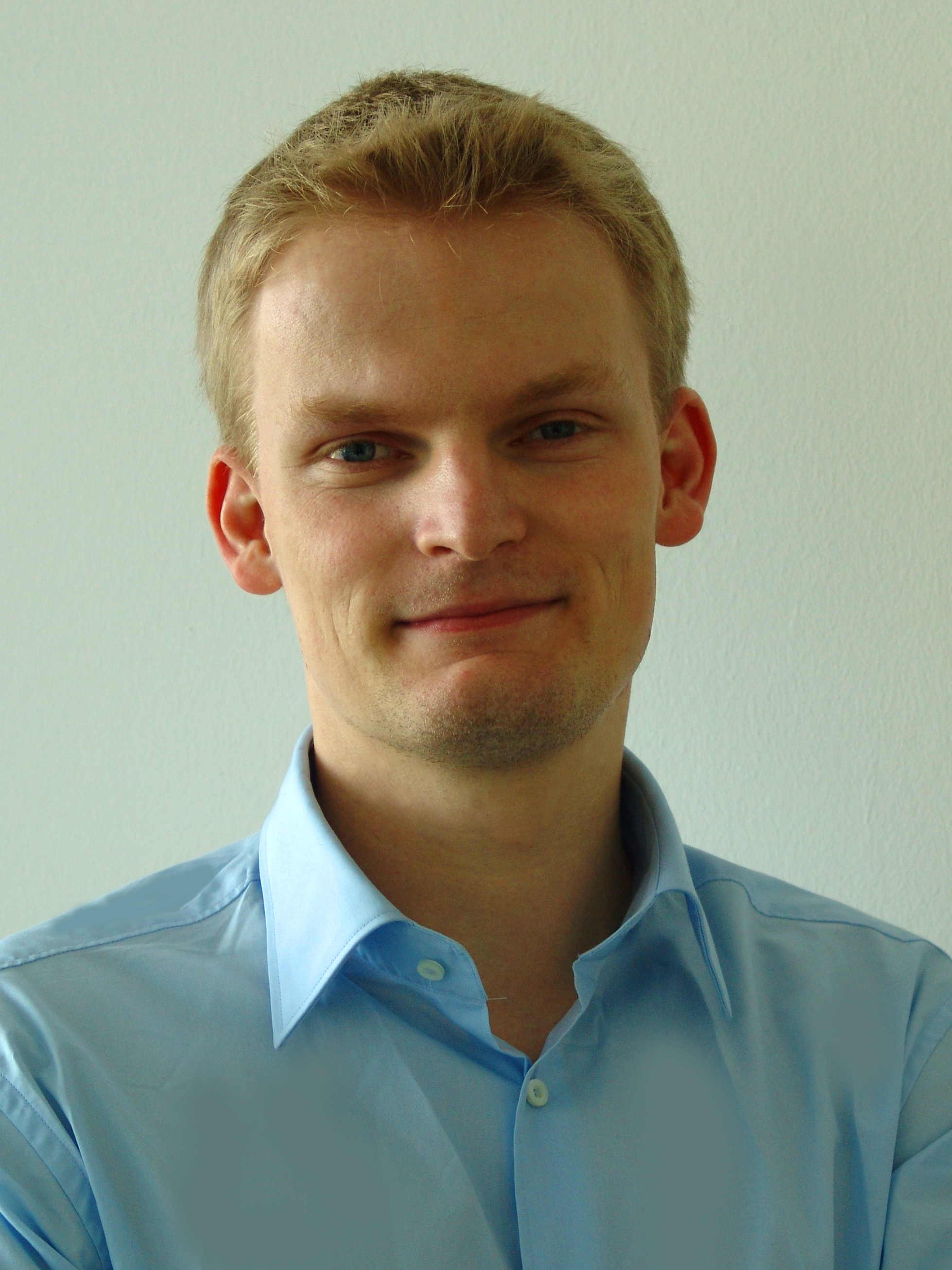}}]{Robert Wille}
Robert Wille (M’06–SM’09) is Full Professor at the Johannes Kepler University Linz, Austria, and Chief Scientific Officer at the Software Competence Center Hagenberg, Austria. He received the Diploma and Dr.-Ing. degrees in Computer Science from the University of Bremen, Germany, in 2006 and 2009, respectively. Since then, he worked at the University of Bremen, the German Research Center for Artificial Intelligence (DFKI), the University of Applied Science of Bremen, the University of Potsdam, and the Technical University Dresden. Since 2015, he is
working in Linz/Hagenberg. His research interests are in the design of circuits and systems for both conventional and emerging technologies. In these areas, he published more than 300 papers in journals and conferences and served in editorial boards and program committees of numerous journals/conferences such as TCAD, ASP-DAC, DAC, DATE, and ICCAD. For his research, he was awarded, e.g., with a Best Paper Award at ICCAD, a DAC Under-40 Innovator Award, a Google Research Award, and more.
\end{IEEEbiography}

\end{document}